\documentclass[prd,onecolumn,nofootinbib,showpacs,showkeys]{revtex4}
\usepackage{latexsym,graphicx,amssymb,amsmath,mathrsfs} 
\usepackage{setspace,bm}

\newcommand{\be}{\begin{equation}}    
\newcommand{\ee}{\end{equation}}    
\newcommand{\bea}{\begin{eqnarray}}    
\newcommand{\eea}{\end{eqnarray}}    
    
\newcommand{\rt}[1]{{}}    

\newcommand{\un}{\mbox{$\mathbf{1}$ \hspace{-0.91em}
    \raisebox{0.05em}[0pt]{$\shortmid$} \hspace{-0.895em} \raisebox{0.235em}[0pt]{$\shortmid$}}}
 
\makeatletter
\renewcommand\appendix{\par
  \setcounter{section}{0}%
  \setcounter{subsection}{0}%
  \gdef\thesection{\appendixname\space\@Alph\c@section}}
\makeatother

\begin{document}    
{\allowdisplaybreaks

\title{Renormalized effective actions  for the $O(N)$ model at
  next-to-leading order of the $1/N$ expansion}

\author{G. Fej\H{o}s}
\email{geg@ludens.elte.hu}
\affiliation{Department of Atomic Physics, E{\"o}tv{\"o}s University,
H-1117 Budapest, Hungary}

\author{A. Patk{\'o}s}
\email{patkos@galaxy.elte.hu}
\affiliation{Department of Atomic Physics, E{\"o}tv{\"o}s University,
H-1117 Budapest, Hungary}

\author{Zs. Sz{\'e}p}
\email{szepzs@achilles.elte.hu}
\affiliation{Research Group for Statistical and Biological Physics 
of the Hungarian Academy of Sciences, H-1117 Budapest, Hungary}

\begin{abstract}
A fully explicit renormalized quantum action functional is constructed
for the $O(N)$-model in the auxiliary field formulation at
next-to-leading order (NLO) of the $1/N$ expansion. Counterterms are
consistently and explicitly derived for arbitrary constant vacuum
expectation value of the scalar and auxiliary fields. The renormalized
NLO pion propagator is exact at this order and satisfies Goldstone's
theorem. Elimination of the auxiliary field sector at the level of the
functional provides with ${\cal O}(N^0)$ accuracy the renormalized 
effective action of the model in terms of the original variables. 
Alternative elimination of the pion and sigma propagators provides 
the renormalized NLO effective potential for the expectation values 
of the $N$-vector and of the auxiliary field with the same accuracy.
\end{abstract}


\maketitle

\section{Introduction}
Large-$N$ expansion is a classical non-perturbative tool of quantum
field theory \cite{dolan74,schnitzer74,coleman74}. Leading order
solution of the $O(N)$ symmetric model has been applied to interesting
problems of finite temperature phenomena, in particular the study of
the restoration of the $SU(2)\times SU(2)\sim O(4)$ chiral symmetry of
QCD \cite{meyer-ortmanns93,bocskarev96,patkos02}. One of its
attractive points is that it preserves Goldstone's theorem at every
order of the expansion, a feature not shared by all resummations of
the original perturbative series.  Being a resummation to all orders,
it has some features which are absent at any given order of the
perturbation theory, but which are believed to be true for the exact
solution. One example is the renormalization scale invariance
observed at leading order (LO) of the $1/N$ expansion.  On the other
hand, related to the now well established triviality of scalar
theories (see {\it e.g.} \cite{kuti} for a review), it shows the presence 
of a tachyonic pole (Landau ghost) in the renormalized propagators.

In early publications \cite{schnitzer74,coleman74,root74} the
appearance of tachyons was considered as an inconsistency of the
large-$N$ approximation.  In next-to-leading order (NLO)
investigations of the expansion, started already in 1974
\cite{root74}, the tachyonic problem seemed to be aggravated because
the renormalized effective potential appeared to be complex for all
values of the field. This led to the claim that the large-$N$
expansion breaks down. Extensive studies of the $O(N)$ symmetric model
\cite{abott76,moshe83} established the consistency of the $1/N$
expansion for the effective potential and revealed its rich phase
structure.  Considered in a restricted sense, as a renormalized
effective theory, the large-$N$ expansion turned out to be a valuable
nonperturbative tool, when applied to phenomena dominated by scales
much lower than the cut-off.  A strict cut-off version of the model
was considered in \cite{schnitzer95} showing that with restrictions on
the value of the background field the model has a phase with
spontaneously broken symmetry, free of tachyons.  A different solution
to the tachyonic problem, called the tachyonic regularization, was
proposed in \cite{binoth98,binoth99a}, where the tachyonic pole is
minimally subtracted. Most of the studies were realized in a
reformulation of the model in which the quartic self-coupling of the
$N$-component scalar field is replaced by an auxiliary field mediated
interaction.

The renormalization of the $O(N)$ model was performed at NLO in
\cite{root74}. While pointing out all the divergent integrals, this
analysis skipped the explicit calculation of the counterterms which
were determined only at LO. Calculation of the self-coupling
counterterm in \cite{smekal94} showed that at NLO the $\beta$-function
is corrected by terms of order $1/N.$ The success of the
renormalization program of the two-particle irreducible (2PI) approximation
\cite{hees02,blaizot04,berges05} revived also the study of the
renormalizability of the NLO corrections of the large-$N$ expansion
\cite{cooper04,cooper05}. Despite all these efforts, the detailed and
explicit knowledge of the counterterms is missing in the literature,
even today. More recent publications with interesting
finite temperature applications raised doubts about the
renormalizability of the NLO approximation for arbitrary values of the
field expectation and away from the saddle point value of the
auxiliary field \cite{andersen04,andersen08}. This issue was addressed
in an original paper by Jakov\'ac \cite{jakovac08}, where a momentum
dependent counterterm is introduced for the auxiliary field, which
keeps its ``propagator'' at its classical expression. This
generalization of the set of allowed counterterms was shown to lead to
renormalizability for arbitrary field and auxiliary field
expectations.

The present contribution provides an explicit construction with 
${\cal O}(N^0)$ accuracy of the
generalized renormalized effective action in the auxiliary
field formulation of the $O(N)$ model. This generalisation considers
beyond the fields also their 2-point functions as dynamical variables
of the action functional. This feature is common with the approach of
2PI ($\Phi-$derivable) formalism. Subsequent eliminations of selected
partial sets of field variables and two-point functions lead to
functionals depending on remaining variables. We start from the 1PI
Dyson-Schwinger (DS) equations of the model in the auxiliary field
formulation. An action functional will be given for that specific 
closure of the infinite set of DS-equations which keeps the coupling of the
auxiliary field at its classical value. This closure condition is
called sometimes the Bare Vertex Approximation (BVA truncation)
\cite{mihaila01}.

The leading order part of the effective action is proportional to the
number of degrees of freedom $N$.  At next-to-leading order (NLO) our
goal is to construct the renormalised effective action with ${\cal
  O}(N^0)$ accuracy in an arbitrary field background. This goal will
be achieved by expanding the propagators in inverse powers of $N$ and
omitting from the action functional all terms which contribute beyond
NLO. Then the actual task is to construct the renormalizing counter
functional for this approximation, which preserves also Goldstone's
theorem, obviously obeyed in the NLO $1/N$-expansion of the bare
theory. We remark here that the 2PI-1/N approximation where one looks
for a self-consistent solution of the stationarity conditions of the
effective action with respect to the propagators represents a
different resummation strategy.  This is reflected also by the fact
that the counter-functional we determine is not applicable in the 2PI
renormalisation program.

The fact that 2-point functions will be determined following the $1/N$
hierarchy and not self-consistently has the definite backward effect
of implying secular behavior in time dependent applications
\cite{mihaila01,yaffe00}. It is, however, essential for ensuring
Goldstone's theorem, since an ${\cal O}(1/N)$ expansion on the level
of the 2PI generating functional was shown to not cure the violation
of Goldstone's theorem by the self-consistent propagator
\cite{cooper05}.  Proposition for the preservation of Goldstone's
theorem for the self-consistent (2PI) propagators exists at present
only at LO (Hartree-level) \cite{ivanov05} and follows efforts
initiated in the framework of non-relativistic many-body theory
\cite{hohenberg65}.

By explicit construction of the counter functional in the auxiliary
field formulation we demonstrate that the model is renormalizable at
NLO in the large-$N$ expansion for arbitrary vacuum expectation values
of sigma and auxiliary fields. Elimination of the auxiliary field and
related propagators at the level of the ${\cal O}(N^0)$ accurate
functional leads to the recovery of ${\cal O}(N)$ and ${\cal O}(N^0)$
terms of the NLO 2PI effective action of the $O(N)$ model
\cite{dominici93}, this time completed with all renormalizing 
counterterms of the same accuracy (valid, however, only 
in case of NLO large $N$ expansion of its propagators). Alternatively, 
elimination of the NLO pion and LO sigma propagators produces the 
renormalized effective potential for the sigma and the auxiliary 
fields \cite{andersen04}, with correct counterterms.

The method of the analysis and separation of the overall and
subdivergences follows very closely our previous work
\cite{fejos08,patkos08} on the counterterm construction to
2PI-functionals in constant field background.  Here, the actual
functional dependence of the divergences on the (auxiliary) field
background dictates the form of the necessary counterterms. The main
issue to be demonstrated is the mutual consistency of the different
counterterms required to cancel the divergences present in different
equations. This can be done the most conveniently with reference to a
unique counterterm functional. Also the elimination of different
subsets of the variables can be realised most efficiently by 
manipulating the expression of the effective renormalised action 
rather than on the level of the field- and propagator-equations.
 
In Section~2 we shortly outline the construction of the effective
  action functional starting from the Dyson-Schwinger equations, by
explicitly formulating the closure condition. In Section~3 the leading
order renormalization will be presented, which illustrates through a
``textbook'' example our logic to be followed at NLO. The main results
of our paper are contained in Section~4, where all pieces of the NLO
counterterm functional are obtained. Here, we rely on some detailed
considerations presented in \ref{sec:Div}. In Section~5 we collect the
pieces of the counterterm functional into a unique expression. The
elimination of the auxiliary field in Section~6 leads to the
renormalized NLO functional written exclusively in terms of the
original fields and their propagators. Alternative elimination of the
propagators of the pion and sigma fields provides us with the NLO
renormalized effective potential as function of the vacuum expectation
values of sigma and auxiliary fields.  Some details of this procedure
are given in \ref{sec:Veff}.  We summarize the extensive material of
the paper in Section~7.

\section{From Dyson-Schwinger equations to the generalised effective action}

At the level of the generating functional $Z[J]$ one introduces 
the auxiliary field $\alpha$ into the $O(N)$ model through the
functional Hubbard-Stratonovich transformation \cite{HST}:
\be
\nonumber
\int[d\alpha] \exp\bigg\{i\int d^4 x\bigg[
-\frac{1}{2}\alpha^2(x)+\frac{i}{2}\sqrt{\frac{\lambda}{3N}}
\alpha(x)\phi^2(x)\bigg]\bigg\}
\propto
\exp\left\{i\int d^4 x\left[
-\frac{\lambda}{24 N}\big(\phi^2(x)\big)^2
\right]
\right\},
\ee
where $\phi^2=\sigma^2+\pi_n^2.$
Then, the  extended action in constant background $\sqrt{N}v$ reads:
\bea
S[\sigma,\pi_n,\alpha,v]=\int
d^4x\bigg[\frac{1}{2}\big(\partial_\mu\sigma(x)\big)^2
+\frac{1}{2}\big(\partial_\mu\pi_n(x)\big)^2-
\frac{m^2}{2}\Big(\sigma^2(x)+\pi^2_n(x)+
2\sqrt{N}v\sigma(x)+N v^2\Big)
\nonumber\\
-\frac{1}{2}\alpha^2(x)+\frac{i}{2}
\sqrt{\frac{\lambda}{3N}}\alpha(x)
\Big(\sigma^2(x)+\pi_n^2(x)+2\sqrt{N}v\sigma(x)+N v^2\Big)\bigg].
\label{Eq:S_aux}
\eea

The infinite hierarchy of Dyson-Schwinger equations is obtained 
from the master equation \cite{rivers}
\be
-J_A(x)=\frac{\delta \Gamma[\phi_B]}{\delta \phi_A(x)}=
\frac{\delta S(\phi_B)}{\delta \phi_A(x)}\Big|_{\phi_B=\hat\phi_B}\un,
\label{Eq:master}
\ee 
by functional derivation with respect to the field. The notation on
the right hand side means that in the derivative of the action a
generic field $\phi_B(x)$ is replaced by
$\hat\phi_B(x)=\phi_B(x)+\int_y G_{BC}(x,y)\frac{\delta}{\delta \phi_C(y)}$ 
and the resulting expression acts on the unity in the field space. 
$J_A(x)$ is a current coupled to the field $\phi_A(x)$ while 
$\Gamma[\phi_B]$ is the effective action from which the inverse of the 
exact two-point function of two fields $\phi_A$ and $\phi_B$ can be 
derived as $i G_{AB}^{-1}(x,y)=\delta^2\Gamma/\delta\phi_A(x)\delta\phi_B(y).$ 
The index $A$ denotes the ``flavor'' of the field ($\alpha, \sigma$ or
$\pi_n$) and possible $O(N)$ indices. Repeated capital indices mean summation 
over the range of ``flavor'' indices.

Since $J_A$ is the current for which the field expectation value of
$\phi_A$ has a prescribed value, if one shifts the auxiliary field
$\alpha(x)$ by its rescaled ``spontaneous'' expectation value, that is
$\alpha(x)\to\alpha(x)+\hat\alpha\sqrt{3N/\lambda},$ then in
(\ref{Eq:master}) one has to take $J_A\to 0$ for
$A\in\{\sigma,\alpha\}.$ One obtains in this way the quantum equation
of state for $v$ and the saddle point equation for~$\hat\alpha:$
\bea
&&N v\left[-m^2+i\hat\alpha\right]+i
\sqrt{\frac{\lambda}{3}}G_{\alpha\sigma}(x,x)=0,
\nonumber
\\
&&-\frac{3N}{\lambda}\hat\alpha+\frac{i}{2}
\left[N v^2+(N-1)G_\pi(x,x)+G_{\sigma\sigma}(x,x)
\right]=0.
\label{Eq:DS1}
\eea
Here, $G_\pi, G_{\sigma\sigma}$ and $G_{\alpha\sigma}$ are components
of the exact propagator matrix defined in the ``flavor'' space
spanned by the fields $\alpha, \sigma, \pi_n$. (For the sake of
brevity double indices are used only for the mixing components of the
propagator matrix).

The next layer of the Dyson-Schwinger equations can be written for the
propagators by differentiating (\ref{Eq:master}) with respect to the
field and setting in the end $J_A\to 0.$ One obtains:
\bea
i(G^{-1})_{\sigma\sigma}(x,y)&=&i(D^{-1})_{\sigma\sigma}(x,y)-
\sqrt{\frac{\lambda}{3N}}\int_z\int_w G_{\alpha A}(x,z)G_{\sigma B}(x,w)
\Gamma_{A B\sigma}(z,w,y),\nonumber\\ 
i(G^{-1})_{\sigma\alpha}(x,y)&=&i(D^{-1})_{\sigma\alpha}(x,y)-
\sqrt{\frac{\lambda}{3N}}\int_z\int_w G_{\alpha A}(x,z)G_{\sigma B}(x,w)
\Gamma_{A B\alpha}(z,w,y),\nonumber\\ 
i(G^{-1})_{\alpha\alpha}(x,y)&=&i(D^{-1})_{\alpha\alpha}(x,y)-\frac{1}{2}
\sqrt{\frac{\lambda}{3N}}\int_z\int_w\Big[G_{\sigma A}(x,z)G_{\sigma B}(x,w)
+G_{\pi_n A}(x,z)G_{\pi_m B}(x,w)\Big]\Gamma_{A B\alpha}(z,w,y),\ \ \nonumber\\
i(G^{-1})_{\pi_n\pi_m}(x,y)&=&i\delta_{n m}D^{-1}_\pi(x,y)-
\sqrt{\frac{\lambda}{3N}}\int_z\int_w G_{\alpha A}(x,z)G_{\pi_n B}(x,w)
\Gamma_{A B\pi_m}(z,w,y).
\label{Eq:DS2}
\eea
The tree-level propagators appearing here have in Fourier space the
following expressions:
\be
i(D^{-1})_{\sigma\sigma}(k)=i D^{-1}_\pi(k)=k^2-M^2,
\qquad i(D^{-1})_{\alpha\alpha}(k)=-1,
\qquad i(D^{-1})_{\alpha\sigma}(k)=i\sqrt{\frac{\lambda}{3}}v,
\label{Eq:tree_prop}
\ee
where we introduced the shorthand notation
\[
M^2=m^2-i\hat\alpha.
\]

Exact 3-point vertices denoted by $\Gamma_{A B C}(x,y,z)=
\delta^3\Gamma/\delta\phi_A(x)\delta\phi_B(y)\delta\phi_C(z)$ also 
appear in these equations. The infinite set of DS-equations can be 
closed in a simple
way still treating the one- and two-point functions dynamically by 
setting for these vertex functions their tree-level (classical) 
expressions:
\be
\Gamma_{\sigma\sigma\alpha}(x,y,z)=i\sqrt{\frac{\lambda}{3N}}
\delta(x-y)\delta(x-z), \qquad 
\Gamma_{\pi_n\pi_m\alpha}(x,y,z)=\delta_{n m}
\Gamma_{\sigma\sigma\alpha}(x,y,z).
\label{Eq:vertices}
\ee
For any other set of indices the 3-point vertex vanishes at classical level.

With this closure of the Dyson-Schwinger hierarchy, the set of
equations given in (\ref{Eq:DS1}) and (\ref{Eq:DS2}) can be derived
upon variation with respect to the corresponding quantities from the
following multivariable generalised effective action:
\bea
\Gamma[\hat\alpha,v,G_\pi,{\cal G}]&=&\frac{1}{2}\left
(m^2-i\hat\alpha\right)N v^2+\frac{3N}{2\lambda}\hat\alpha^2
\nonumber\\
&-&\frac{i}{2}\int_k\Biggl[
(N-1) \left(\ln G^{-1}_\pi(k) + D^{-1}_\pi(k)G_\pi(k)\right)
+\textrm{Tr}\ln {\cal G}^{-1}(k)
+\textrm{Tr}\left( {\cal D}^{-1}(k) {\cal G}(k)\right)
\Biggr]
\nonumber\\
&+&i\frac{\lambda}{12N}\int_k\int_p\Biggl[G_{\alpha\alpha}(k)
\Big((N-1)G_\pi(p)G_\pi(p+k)
+G_{\sigma\sigma}(p)G_{\sigma\sigma}(p+k)\Big)\nonumber\\
&&+2G_{\alpha\sigma}(p)G_{\sigma\sigma}(k)G_{\sigma\alpha}(p+k)\Biggr]
+\Delta\Gamma[\hat\alpha,v,G_\pi,{\cal G}].
\label{Phi-funct}
\eea
Here, $m^2$ and $\lambda$ are the renormalized couplings, and
$\Delta\Gamma[\hat\alpha,v,G_\pi,{\cal G}]$ is the yet undetermined
counterterm functional.  ${\cal G}$ and ${\cal D}$ are two symmetric
$2\times 2$ matrices with components
$G_{\sigma\sigma},G_{\sigma\alpha},$ $G_{\alpha\alpha}$ and
$D_{\sigma\sigma},D_{\sigma\alpha},$ $D_{\alpha\alpha},$ respectively.
Their inverse matrices are denoted by ${\cal G}^{-1}$ and 
${\cal D}^{-1}$, respectively.

The functional in (\ref{Phi-funct}) corresponds to a specific
 $\Phi$-derivable action.  Except for the last term in the
square bracket of the last integral, it reproduces pieces of the NLO
2PI effective action presented in (44) and (55) of \cite{aarts02}.
This additional term, which comes from the self-energy of the second
equation in (\ref{Eq:DS2}), is catalogued as next-to-NLO (NNLO) 
in \cite{aarts02} and corresponds to Fig.~7a of that reference.

Having in mind the ${\cal O}(N^0)$ determination of the effective action 
we realize that most terms of the square bracket of the last integral in 
(\ref{Phi-funct}) contribute only at NNLO. Therefore we truncate further 
the functional in (\ref{Phi-funct}) and will work with the approximate 
effective action
\bea
\Gamma[\hat\alpha,v,G_\pi,{\cal G}]&=&\frac{1}{2}\left
(m^2-i\hat\alpha\right)N v^2+\frac{3N}{2\lambda}\hat\alpha^2
\nonumber\\
&-&\frac{i}{2}\int_k\Biggl[
(N-1) \left(\ln G^{-1}_\pi(k) + D^{-1}_\pi(k)G_\pi(k)\right)
+\textrm{Tr}\ln {\cal G}^{-1}(k)
+\textrm{Tr}\left( {\cal D}^{-1}(k) {\cal G}(k)\right)
\Biggr]
\nonumber\\
&+&i\frac{\lambda}{12}\int_k\int_pG_{\alpha\alpha}(k)
G_\pi(p)G_\pi(p+k)+\Delta\Gamma[\hat\alpha,v,G_\pi,{\cal G}].
\label{Gamma-funct}
\eea
This truncated form of the functional influences the propagator
equations of the $\alpha - \sigma$ sector (at LO) and of the pion
fields (at NLO). We emphasize, that even if one preserved the
omitted terms, the ${\cal O}(1/N)$ solution of that approximation would
not correspond to the full $1/N$-accurate solution of the
Dyson-Schwinger equations in the $\alpha - \sigma$ sector. This is a
consequence of the closure (\ref{Eq:vertices}), which does not take
into account that the vertex $\Gamma_{\alpha\alpha\alpha}$ starts contributing 
at order $1/\sqrt{N}$ through a one-loop diagram in which  
the pion multiplicity compensates the suppression resulting from having three
$\Gamma_{\pi\alpha\alpha}$ vertices. This results in contributions of 
${\cal O}(1/N)$ to both $i(G^{-1})_{\sigma\alpha}$ and $i(G^{-1})_{\alpha\alpha}$
through the terms $G_{\alpha \alpha}G_{\sigma\alpha}\Gamma_{\alpha\alpha\alpha}$ 
and $G_{\sigma \alpha} G_{\sigma \alpha}\Gamma_{\alpha\alpha\alpha}$ of the 
second and third equations of  (\ref{Eq:DS2}).

Renormalizability of this approximation will be investigated at the
level of the derivatives of (\ref{Gamma-funct}) up to the
next-to-leading order of the large-$N$ expansion of the equations for
the pion propagator and the background fields. One attempts the
construction of an appropriate counterterm functional
$\Delta\Gamma[\hat\alpha,v,G_\pi,{\cal G}]$. This countertem functional 
allows for a uniform treatement of the counterterms and also makes possible 
to keep track of the effect a counterterm determined from one equation 
has in the renormalization of another equation. Some new insight will be
offered when compared to other approaches, where the propagators are
not considered as variational variables \cite{andersen04,andersen08},
or when the quantities related to the auxiliary field are eliminated
\cite{dominici93}. We shall demonstrate the validity of Goldstone's
theorem also for the renormalized NLO propagators.  We start our
program with the short description of the leading order analysis.

\section{Leading order (LO) construction of the counterterms}

\subsection{Saddle point equation (SPE)}

Taking the derivative of the functional in (\ref{Gamma-funct}) 
with respect to $\hat\alpha$ one arrives at the expression:
\be
\frac{\delta \Gamma}{\delta\hat\alpha}[\hat\alpha,v,G_\pi,{\cal G}]=
\frac{3N}{\lambda}\hat\alpha-i\frac{N}{2}
\left(v^2+\int_k D_\pi(k)\right)+{\textrm c.t.},
\label{Eq:LO-SPE}
\ee
where we replaced $G_\pi,$ originally appearing in the integral above,
by $D_\pi$ introduced in (\ref{Eq:tree_prop}). The last term denoted by
``c.t.'' is the
contribution of the counterterm functional which has to be
constructed to ensure the finiteness of this equation. The structure 
of divergences for the tadpole integral above is given in \ref{sec:Div}.
Using (\ref{Eq:tadpole}), one can see that the finiteness of
(\ref{Eq:LO-SPE}) for {\it any} value of $\hat\alpha$ is ensured by
the following ${\cal O}(N)$ counterterm functional:
\be
\Delta\Gamma^{\alpha,N}=i\hat\alpha\frac{N}{2}\left[T_d^{(2)}+\left(m^2-
M_0^2-i\frac{1}{2}\hat\alpha\right)T_d^{(0)}\right].
\label{Eq:ct_aN}
\ee
Here, $T_d^{(2)}$ and $T_d^{(0)}$ are 
the quadratic and logarithmic divergences of the pion tadpole as 
given in (\ref{Eq:Td2}) and (\ref{Eq:Td0}) of \ref{sec:Div}.
Then, from (\ref{Eq:LO-SPE}) one obtains the finite saddle point equation
\be
\frac{3N}{\lambda}\hat\alpha-i\frac{N}{2}\left(v^2+T_{\pi}^F\right)=0,
\ee
where $T_{\pi}^F$ is the finite part of the pion tadpole integral.

\subsection{Equation of state and Goldstone's theorem}

The leading order term in the derivative of the
functional $\Gamma$ with respect to $v$ is of order $N$
\be
\frac{\delta \Gamma}{\delta v}[\hat\alpha,v,G_\pi,{\cal G}]=
N v M^2.
\label{LO-EOS}
\ee
The right hand side is finite in itself, it does not necessitate the
introduction of any ${\cal O}(N)$ counterterm.

Since the equation of state is obtained by equating the r.h.s. of
(\ref{LO-EOS}) to zero and the leading order inverse pion
propagator is of the form (\ref{Eq:tree_prop}), one immediately sees that
Goldstone's theorem ($D_\pi^{-1}(\hbox{$k=0$})=0$) is obeyed.

\subsection{Leading order propagator matrix in the $\alpha - \sigma$-sector}
 
The only entry of the $2\times 2$ inverse propagator matrix which receives
LO correction is $(G^{-1})_{\alpha\alpha}:$
\be
i(G^{-1})_{\alpha\alpha}(k)=-1+\frac{\lambda}{6}I_\pi^F(k)+
\frac{\lambda}{6}T_d^{(0)}+c.t.\,.
\ee
The definitions of $I_\pi$ and of its finite part $I_\pi^F$ in terms of $D_\pi,$
which at LO replaces $G_\pi,$ are given in \ref{sec:Div} (see 
(\ref{Eq:mid1})).  In order to make
this equation finite one has to introduce another 
piece into the counterterm functional:
\be
\Delta\Gamma^{\alpha\alpha}=\frac{\lambda}{12}T_d^{(0)}
\int_k G_{\alpha\alpha}(k).
\label{phi-alpha-alpha}
\ee
The LO matrix elements of the $2\times 2$ propagator matrix are then
\be
G_{\sigma\sigma}^{(0)}(k)=\left(1-\frac{\lambda}{6} I_\pi^F(k)\right) 
\tilde G(k),\qquad
G_{\alpha\alpha}^{(0)}(k)=-i D_\pi^{-1}(k)\tilde G(k),\qquad
G_{\alpha\sigma}^{(0)}(k)=i \sqrt{\frac{\lambda}{3}}v\tilde G(k),
\label{Eq:G_LO_matrix}
\ee
where 
\be
i\tilde G^{-1}(k)=(k^2-M^2)\left(1-\frac{\lambda}{6}
I_\pi^F(k)\right) -\frac{\lambda}{3} v^2.
\label{Eq:Gt}
\ee
In the broken symmetry phase all elements of the LO propagator matrix
have common pole structure determined by $\tilde G(k)$.  This is the
manifestation of the hybridization for $v\ne 0$ of the longitudinal
field component $\sigma$ and of the composite field $\alpha\sim
\sigma^2+\pi_n^2$ \cite{szepfalusy1974}. This feature is relevant
when studying dynamical aspects of the phase transition at finite
temperature.

Using the first entry of (\ref{Eq:G_LO_matrix}) one can derive 
the following relation between the LO sigma and pion propagators: 
\be
G_{\sigma\sigma}^{(0)}(k)=D_\pi(k)-i\frac{\lambda}{3}v^2
\frac{G_{\sigma\sigma}^{(0)}(k) D_\pi(k)}{1-\lambda I_\pi^F(k)/6}.
\label{Eq:LO_relation}
\ee
This relation will be very useful for the divergence analysis at NLO.

\section{Next-to-leading order (NLO) construction of $\Delta\Gamma$}

The construction of the NLO counterterm starts with discussing the
$1/N$ expansion of the pion propagator. Its asymptotics will be
completely determined by certain integrals of the LO propagators. We
shall see that the counterterm functional can be determined by the
asymptotics of the LO propagators and of the NLO self-energy of the
pions.  In order to explicitly demonstrate the NLO renormalizability
for all values of $v$ and $\hat\alpha$ we investigate the derivatives
of the functional (\ref{Gamma-funct}) with respect to these
variables. The mutual consistency of the counterterms renormalizing
these three equations is fundamental for the outcome of our analysis.

\subsection{Pion propagator, equation of state and Goldstone's theorem}

The pion propagator at NLO in the $1/N$ expansion is given by 
\be
i G^{-1}_\pi(k)=i D^{-1}_\pi(k)-i\frac{\lambda}{3N}\int_p
G_{\alpha\alpha}^{(0)}(p)D_\pi(p+k)+{c.t.}~.
\label{Eq:Gp_NLO}
\ee
Since we need the pion self-energy to ${\cal O}(1/N)$ accuracy, 
we replaced $G_{\alpha\alpha}$ with $G_{\alpha\alpha}^{(0)}$
and $G_\pi$ with $D_\pi$ in the above integral. 

In order to determine the counterterm contribution in
(\ref{Eq:Gp_NLO}) one has to study the divergence of the NLO
self-energy. As shown in \ref{sec:Div} this divergence is momentum
independent, i.e. there is no need for infinite wave function
renormalization, in accordance with \cite{binoth98}.  Using the first
two entries of (\ref{Eq:G_LO_matrix}) together with 
(\ref{Eq:LO_relation}) one can write
\be
G^{(0)}_{\alpha\alpha}(p)=-\frac{i}{1-\lambda I_\pi^F(p)/6}
-\frac{\lambda v^2}{3} \frac{G_{\sigma\sigma}^{(0)}(p)}
{\left(1-\lambda I_\pi^F(p)/6\right)^2}.
\label{Eq:noZ1}
\ee
Then, one sees that the momentum independent divergence is determined by
the first term of (\ref{Eq:noZ1}):
\be
i\int_p G_{\alpha\alpha}^{(0)}(p) D_\pi(p)\bigg|_{\textnormal{div}}=
\int_p\frac{D_\pi(p)}{1-\lambda I_\pi^F(p)/6}
\bigg|_{\textnormal{div}}=:\tilde T_\textnormal{div}(M^2).
\label{pion-p0}
\ee
This divergence is worked out explicitly in \ref{sec:Div} with the result
\be
\tilde T_\textnormal{div}(M^2)=T_a^{(2)}-\frac{\lambda}{2}(M^2-M_0^2)T_a^{(I)},
\label{tilde-div}
\ee
where the quadratically and logarithmically divergent integrals 
$T_a^{(2)}$ and $T_a^{(I)}$ are defined in (\ref{Eq:Ta2I}). 

The second term of (\ref{Eq:noZ1}) does not contribute to the
divergence of the first integral of (\ref{pion-p0}), since iterating
(\ref{Eq:LO_relation}) once one recognizes that by logarithmic(!) power
counting the following integral is actually convergent:
\be
\int_p\frac{D_\pi^2(p)}{\big(1-\lambda I_\pi^F(p)/6\big)^2}.
\ee
Due to this fact we do not encounter any divergence proportional to
$v^2$ in the NLO pion self-energy. 

It is instructive to point out here a peculiarity of the resummation
procedure as compared to the order-by-order perturbative
renormalization.  Namely, when in the above integral the denominator
is expanded in powers of $\lambda$ then at $n$th order of the
expansion one finds a $\lambda^n (\log \Lambda)^{n+1}$ type
divergence.  Through formal resummation of this divergent series a
finite result is obtained. This argument explicitly shows that in a
resummed perturbation theory the structure of the counterterms can be
different from that seen at any given order of the perturbation
theory. The same effect was noticed in \cite{binoth98} in connection
with the wave function renormalization constants of pion and sigma
fields which arise from imposing renormalization conditions on the
residua of their propagators. At NLO in the $1/N$ expansion they are
finite whereas in an expansion to any given order in the coupling they
appear divergent. Similar consideration applies to the divergence
appearing in (\ref{pion-p0}). Expanding the denominator of the middle
expression into powers of $\lambda$ one finds the usual quadratic and
logarithmic divergences characteristic for the perturbative
contributions. Resummation modifies these divergences as can be 
explicitly seen in the last two formulas of \ref{sec:Div}.

Since the necessary counterterm in (\ref{Eq:Gp_NLO}) compensates
$\tilde T_\textnormal{div},$ one readily finds the required counterterm
functional piece upon functional integration of (\ref{tilde-div}) with
respect to $G_\pi$ and multiplying it by $\lambda/6$:
\be
\Delta\Gamma^{\pi}=-
\frac{\lambda}{6}
\left[T_a^{(2)}-\frac{\lambda}{2}(M^2-M_0^2)
T_a^{(I)}\right]\int_k G_\pi(k).
\label{Eq:ct_Gp}
\ee

Next, one investigates the renormalization of the derivative of the
2PI effective action with respect to the background $v.$ At NLO in the
$1/N$ expansion this is given by
\bea
\frac{\delta \Gamma}{\delta v}[\hat\alpha,v,G_\pi,{\cal G}]
&=&
N v M^2-i\sqrt{\frac{\lambda}{3}}\int_k G_{\alpha\sigma}^{(0)}(k)
+c.t.
\nonumber
\\
&=&N v\left[M^2+\frac{\lambda}{3N}\int_k \tilde G(k)\right]+c.t.\,,
\label{Eq:v_derivalt}
\eea
where for the second equality we used the last entry of
(\ref{Eq:G_LO_matrix}). The counterterm functional $\Delta\Gamma^{\pi}$
determined above does not contribute since its derivative 
with respect to $v$ is zero.

The equation of state is obtained by equating the r.h.s. of
(\ref{Eq:v_derivalt}) to zero. Its unrenormalized expression obviously
implies when confronted with (\ref{Eq:Gp_NLO}) the validity of
Goldstone's theorem with ${\cal O}(1/N)$ accuracy.  Note that, as it
is well known, Goldstone's theorem is not followed if one proceeds in
strict 2PI sense which requires the self-consistent determination of
the full propagator without expansion in $1/N.$

In a renormalization procedure which preserves Goldstone's theorem one
should construct a counterterm which does not depend on $G_\pi,$
therefore does not interfere with its already renormalized equation.
Since the divergence in (\ref{Eq:v_derivalt}) coincides with the
divergence of the NLO pion self-energy, the new contribution to the
counterterm functional is obtained upon integrating with respect to
$v$ the expression given in (\ref{tilde-div}) multiplied by 
$\lambda v/3.$ One finds
\be
\Delta\Gamma^{v}=-\frac{\lambda}{6}v^2\left[
T_a^{(2)}-\frac{\lambda}{2}(M^2-M_0^2)T_a^{(I)}\right].
\label{Eq:ct_v}
\ee

We conclude this part by giving the finite pion propagator at NLO in
the $1/N$ expansion, including also the contribution of the
counter-functional $\Delta\Gamma^{\pi}.$ It reads as
\be
i G^{-1}_\pi(k)=k^2-M^2-\frac{\lambda}{3N}\Sigma^{F}_\pi(k),
\qquad 
\Sigma^{F}_\pi(k)=i\int_p G_{\alpha\alpha}^{(0)}(p) D_\pi(k+p)
-\tilde T_\textnormal{div}(M^2).
\label{pi-prop-1}
\ee 
A remarkable feature of this ${\cal O}(1/N)$ renormalized solution is
that it satisfies Goldstone's theorem for arbitrary values of
$\hat\alpha$~!

\subsection{Saddle point equation}

In writing down the derivative of the effective potential with respect
to $\hat\alpha$ one has not to forget about the contributions of the
$\hat\alpha$-dependent counterterms $\Delta\Gamma^{\alpha,N},$
$\Delta\Gamma^{\pi},$ and $\Delta\Gamma^{v}$ constructed above (see
(\ref{Eq:ct_aN}), (\ref{Eq:ct_Gp}), and (\ref{Eq:ct_v})):
\bea
\frac{\delta\Gamma}{\delta \hat\alpha}[\hat\alpha,v,G_\pi,{\cal G}]
&=&
\frac{3N}{\lambda}\hat\alpha-i\frac{N}{2}
\left(v^2+\int_k G_\pi(k)\right)-\frac{i}{2}
\int_k \Big(G_{\sigma\sigma}(k)-G_\pi(k)\Big)\nonumber\\
&&+i\frac{N}{2}\left[T_d^{(2)}+(M^2-M_0^2)T_d^{(0)}\right]
-i\frac{\lambda^2}{12}T_a^{(I)}\left(v^2+\int_k G_\pi(k)\right)
+\frac{\delta\Delta\Gamma^{\alpha, 0}(\hat\alpha)}{\delta\hat\alpha}.
\label{counter-spe-div}
\eea
The contribution of the counterterms determined by the renormalization
of the equation of the inverse pion propagator  (\ref{Eq:Gp_NLO}) and
of the equation of state is displayed in the last but one term. 
\rt{Such type of ``counterterm cross-contribution'' would be more
difficult to recognize on the level of the Dyson-Schwinger equations.}

The last, yet undetermined part of the NLO counterterm functional, 
{\it i.e.} $\Delta\Gamma^{\alpha, 0},$ provides the NLO completion to
$\Delta\Gamma^{\alpha, N}$.  In order to determine it one first has to
evaluate  with NLO accuracy the pion tadpole in the second term of 
the r.h.s. of (\ref{counter-spe-div}). Taking the inverse of $G_\pi^{-1}$
given in (\ref{pi-prop-1}) and expanding it to ${\cal O}(1/N)$ one obtains
\bea
\int_k G_\pi(k)
&=&\int_k D_\pi(k)+\frac{\lambda}{3N}\int_k D_\pi^2(k)
\int_p G_{\alpha\alpha}^{(0)}(p) D_\pi(k+p) + \frac{\lambda}{3N}
i \tilde T_\textnormal{div}(M^2) \int_k D_\pi^2(k)
\nonumber
\\
&=&\int_k D_\pi(k)-i\frac{\lambda}{3N}\left[
\tilde J(M^2)-\tilde T_\textnormal{div}(M^2)\int_k D_\pi^2(k)
\right]
-\frac{\lambda^2 v^2}{9N} J(M^2),
\label{pion-tadpole}
\eea
where for the second equality we used (\ref{Eq:noZ1}) and introduced the 
following functions
\bea
\tilde J(M^2)=\int_k D_\pi^2(k)\int_p 
\frac{D_\pi(p+k)}{1-\lambda I_\pi^F(p)/6},\qquad
J(M^2)=\int_k D_\pi^2(k)
\int_p\frac{D_\pi(p+k)G_{\sigma\sigma}^{(0)}(p)}{(1-\lambda I_\pi^F(p)/6)^2}.
\label{Eq:J_ints}
\eea

Collecting all ${\cal O}(N^0)$ (NLO) divergent terms in 
(\ref{counter-spe-div}) one sees that 
$\delta\Delta\Gamma^{\alpha,0}(\hat\alpha)/\delta\hat\alpha$ is determined by
\bea
\frac{\delta\Delta\Gamma^{\alpha, 0}(\hat\alpha)}{\delta\hat\alpha}&=&
\frac{\lambda}{6}\left[\tilde J_\textnormal{div}(M^2)
-\tilde T_\textnormal{div}(M^2)\int_k D_\pi^2(k)
\right]
-i\frac{\lambda^2}{18}v^2 J_\textnormal{div}(M^2)
\nonumber
\\
&&+\frac{i}{2}\int_k 
\Big(G_{\sigma\sigma}^{(0)}(k)-D_\pi(k)\Big)\bigg|_\textnormal{div}
+i\frac{\lambda^2}{12}T_a^{(I)}\left(v^2+\int_k D_\pi(k)\right),
\label{Eq:Act1}
\eea
where $\tilde J_\textnormal{div}(M^2)$ and $J_\textnormal{div}(M^2)$
denote the divergences of the integrals defined in (\ref{Eq:J_ints}).
Note that to the order of interest it was
again allowed to replace in the last two terms the original 
$G_\pi$ and $G_{\sigma\sigma}$ by $D_\pi$ and $G_{\sigma\sigma}^{(0)},$
respectively.

The important question of consistency inquires whether the previously
constructed counterterms cancel all subdivergences of $\tilde J(M^2)$
and the $v^2$-de\-pen\-dent divergence of second and third terms in
(\ref{Eq:Act1}). The double integral $J(M^2)$ has only overall
divergence, since both $k$ and $p$ integrals are individually
finite. This divergence is determined in (\ref{Eq:v2_div4}) of
\ref{sec:Div}. The divergence of the third term of (\ref{Eq:Act1}) is
determined in (\ref{Eq:v2_div2}) and also shown to be proportional to
$v^2$. One finds that the sum of these two $v^2$-dependent divergences
is canceled by the $v^2$-dependent counterterm contribution appearing
in the last term of (\ref{Eq:Act1}).

The detailed $T=0$ analysis of the divergence structure of $\tilde
J_\textnormal{div}(M^2)$ given in \ref{sec:Div} is based upon the 
explicit expressions of some integrals. The presence of subdivergences
is reflected by the appearance of divergent terms proportional to
$\ln(M^2/M_0^2).$ These should cancel if the approximation is
renormalizable, since in this case only divergences proportional to
powers of $\hat\alpha$ (that is $M^2$) are allowed. The result given
in (\ref{Eq:odiv2}) for $\tilde J_\textnormal{div}(M^2)$ shows when combined
with the second term of the square bracket of (\ref{Eq:Act1}) 
the cancellation of the subdivergence 
$\tilde T_\textnormal{div}(M^2) I_\pi(p=0).$ 
The cancellation of this self-energy type subdivergence of the double
integral $\tilde J(M^2)$ is expected in view of (\ref{pi-prop-1}).

\begin{figure}[!t]
\centerline{ 
\includegraphics[keepaspectratio,width=0.3\textwidth,angle=0]{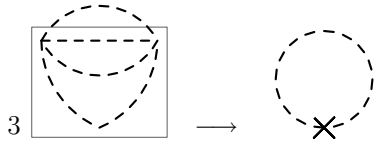}}
\caption{
The appearance of vertex type subdivergences in the first 
double integral of (\ref{Eq:J_ints}) as illustrated 
at leading $\lambda$ 
order in the expansion of the denominator of the integrand. 
The factor of three indicates the possible ways in which 
two lines of the setting-sun diagram form the bubble which corrects 
the vertex at 2-loop level. The cross denotes the associated
lowest order counterterm.
}
\label{Fig:vertex}
\end{figure}

The double integral $\tilde J_\textnormal{div}(M^2)$ has also a
vertex type subdivergence as illustrated in Fig.~\ref{Fig:vertex} at
leading order of the expansion in $\lambda.$ This is canceled, as it
should, by the last integral of (\ref{Eq:Act1}). One can see this
analytically by separating in the difference of the two terms of the
square bracket of (\ref{Eq:Act1}) a contribution proportional to
$T_a^{(I)}T_\pi^F$ which on its turn cancels against the contribution
of the last tadpole integral in (\ref{Eq:Act1}).

With all subdivergences and $v^2$-dependent divergences of
(\ref{Eq:Act1}) canceled, 
$\delta\Delta\Gamma^{\alpha, 0}(\hat\alpha)/\delta\hat\alpha$ 
is determined by the overall divergence of $\tilde J(M^2)$ 
and that of the last tadpole integral. Its expression reads:
\bea
\frac{\delta\Delta\Gamma^{\alpha, 0}(\hat\alpha)}{\delta\hat\alpha}&=&
i T_d^{(2)}+i(4M^2-M_0^2)T_d^{(0)}-i T_a^{(2)}
\left(1+\frac{\lambda}{6}T_d^{(0)}+\frac{\lambda}{48\pi^2}\right)
\nonumber
\\
&&+i T_a^{(I)}\left[\lambda\left(M^2-\frac{1}{2}M_0^2\right)
+\frac{\lambda^2}{6} T_d^{(0)}\left(M^2-M_0^2\right)
+\frac{\lambda^2}{12} T_d^{(2)}+\frac{\lambda^2}{12}\frac{3M^2-M_0^2}{16\pi^2}
\right].
\label{Eq:Act2}
\eea
Since the above expression depends only on $\hat\alpha$ it will have
no ``back-reaction'' neither on the propagator equations nor on the
derivative of the effective potential with respect to the
background. The appearance of terms proportional to $1/(16\pi^2)$
reflects some allowed arbitrariness of the subtraction procedure.

We do not present here the renormalisation of the partial NLO
corrections which would occur in the $\alpha-\sigma$ sector should one
use the effective action functional (\ref{Phi-funct}). All
counterterms necessary for the renormalisation of these pieces are of
${\cal O}(1/N)$, not contributing to the renormalised effective action
at NLO in the $1/N$ expansion.

In conclusion, only divergences proportional to zeroth or first powers
of $M^2$ remained which upon integration over $\hat\alpha$ determine
the $\hat\alpha$-dependent counterterm functional
$\Delta\Gamma^{\alpha,0}(\hat\alpha).$ The counterterms induced by the
renormalization of the NLO pion propagator and of $\delta
\Gamma/\delta v$ played an essential role in the cancellation of
subdivergences and of $v^2$-dependent divergences present in the
expression of $\delta \Gamma/\delta\hat\alpha.$ No limitations
whatsoever showed up on the value of $M^2$ and/or $v^2,$ in contrast
to the findings in \cite{andersen04,andersen08}. We shall return to
the discussion of this difference after reducing the effective action
to the effective potential depending only on the background fields in
\ref{sec:Veff}.  The functional corresponding to the NLO
$1/N$-expansion of the classical vertex approximation (BVA) is
renormalized without any constraint. Our result shows that this can be
achieved also without introducing unconventional counterterms
\cite{jakovac08}.

\section{The explicit form of the counterterm functional}

In this section we collect into a unique expression $\Delta\Gamma$ all
individual pieces determined in Eqs.~(\ref{Eq:ct_aN}),
(\ref{phi-alpha-alpha}), (\ref{Eq:ct_Gp}), (\ref{Eq:ct_v}),
(\ref{Eq:Act2}), and
express it in a conventional form, in which one associates them with
the renormalization of different couplings appearing in the terms
of $\Gamma[\hat\alpha,v,G_\pi,{\cal G}]$ (Eq.~(\ref{Gamma-funct})). The
counterterm functional reads:
\bea
\Delta\Gamma[\hat\alpha,v,G_\pi,{\cal G}]&=&
\frac{1}{2}\left(\delta m^2-i\delta g\hat\alpha\right) v^2+
i\delta\kappa_1\hat\alpha+\delta\kappa_2\hat\alpha^2
\nonumber\\ 
&+&\frac{1}{2}(\delta m^2-i\delta g\hat\alpha)\int_k
G_\pi(k)  
+\frac{1}{2}\delta\kappa_0\int_k G_{\alpha\alpha}(k),
\label{Eq:Phi_ct-funct}
\eea
where the counter-couplings are given by the following expressions:
\bea
\delta m^2&=&-\frac{\lambda}{3}\left[T_a^{(2)}-
\frac{\lambda}{2}\left(m^2-M_0^2\right)T_a^{(I)}\right],\qquad
\delta g=\frac{\lambda^2}{6} T_a^{(I)}, \nonumber\\ 
\qquad
\delta\kappa_2&=&\frac{N+8}{4} T_d^{(0)}+\frac{\lambda}{2} T_a^{(I)}
\left(1+\frac{\lambda}{6} T_d^{(0)}+\frac{\lambda}{64\pi^2}\right),
\qquad
\delta\kappa_0=\frac{\lambda}{6} T_d^{(0)},
\nonumber\\
\delta\kappa_1&=&\frac{N}{2}\left[T_d^{(2)}+(m^2-M_0^2)T_d^{(0)}\right]
+T_d^{(2)}+(4 m^2-M_0^2) T_d^{(0)} 
-T_a^{(2)}\left(1+\frac{\lambda}{6} T_d^{(0)}+\frac{\lambda}{48\pi^2}\right)
\nonumber\\
&&+\lambda T_a^{(I)} \left[
m^2-\frac{1}{2}M_0^2+\frac{\lambda}{12}\left(
2 T_d^{(0)} (m^2-M_0^2) + T_d^{(2)} 
+\frac{1}{16\pi^2}(3m^2-M_0^2)\right)
\right].
\eea 

It is interesting to note that the term proportional to
$\delta\kappa_1$ in (\ref{Eq:Phi_ct-funct}) has no correspondent in
(\ref{Gamma-funct}). Moreover, the terms proportional to
$\delta\kappa_0$ and $\delta\kappa_2$ correspond in
(\ref{Gamma-funct}) to terms not proportional to any coupling of the
original theory. Rather they renormalize numerical coefficients
related to the way the auxiliary field $\alpha$ is introduced by the
parametrization of the Hubbard-Stratonovich transformation
(cf. (\ref{Eq:S_aux})). They correspond to different possibilities of
defining the two-point function of the auxiliary field. They are 
calculated to different orders in $1/N$ since the corresponding 
terms contribute at different levels of the expansion. After 
scaling back the fluctuating part of the field $\hat\alpha$ to $\alpha$ 
the two counter couplings do agree to leading order in $1/N$. This
feature is reminiscent of the renormalisation of the
operators corresponding to different definitions of the n-point
functions in 2PI-approximations. The terms
with $\delta g$ can be regarded as renormalizing the coupling
$g=\sqrt{\lambda/(3N)}$ through which the auxiliary field couples to
fields of the $O(N)$ multiplet. One interprets the renormalisation of
all these operators not occuring in the original model as renormalisations
of the two variants of the 2-point functions of the composite field
$\phi^2$ and of the $\phi^2\sigma\sigma$ vertex (see ch. 30 of 
\cite{zinn-justin02}).

It is notable that $\delta\kappa_2$ determines the renormalization of the
coupling $\lambda$ following the formula
\be
\frac{6}{\lambda_B}=\frac{6}{\lambda}+\frac{4}{N}\delta\kappa_2
=\frac{6}{\lambda}+\frac{N+8}{N}T_d^{(0)}+
2\frac{\lambda}{N} T_a^{(I)}\left(1+\frac{\lambda}{6}T_d^{(0)}+
\frac{\lambda}{64\pi^2}\right).
\label{Eq:lambdaB1}
\ee
This can be seen by looking at the terms proportional to
$\hat\alpha^2$ in the classical part of the functional. This
relation is only intermediary, since it will change in the course of
the elimination of $\hat\alpha$. The part of the counterterm
$\delta\kappa_2$ proportional solely to $T_d^{(0)}$ appears already in
the literature and follows the one-loop $\beta$-function of the model
\cite{kleinert}.  The entire NLO part of $\delta\kappa_2$
proportional to $T_a^{(I)}$ is missing in the analysis of
\cite{andersen04,andersen08} (see e.g. Eq.~(23) of \cite{andersen08}
for the expression of their counterterm $b_1$).

Introducing the following notations
\be
\delta\kappa_2=N\delta\kappa_2^{(0)}+ \delta\kappa_2^{(1)},\qquad
\lambda_B=\lambda+\delta\lambda_\alpha,\qquad
\delta\lambda_\alpha=\delta\lambda_\alpha^{(0)}+
\frac{1}{N} \delta\lambda_\alpha^{(1)},
\label{Eq:lambdaB2}
\ee
one readily obtains
\be
\delta\lambda_\alpha^{(0)}=-\frac{\lambda^2}{6}
\frac{T_d^{(0)}}{1+\lambda T_d^{(0)}/6},\qquad
\lambda_B^{(0)}=\lambda+\delta\lambda_\alpha^{(0)}
=\frac{\lambda}{1+\lambda T_d^{(0)}/6},\qquad
\delta\lambda_\alpha^{(1)}=
-\frac{2}{3}\big(\lambda_B^{(0)}\big)^2 \delta\kappa_2^{(1)}.
\label{Eq:lambdaB3}
\ee
Comparing with Eq.~(25) of \cite{fejos08} one observes that
$\delta\lambda_\alpha^{(0)}$ is the counter-coupling of the $O(N)$
model formulated with its original variables and considered at LO in the
large-$N$ expansion.  Likewise $\lambda_B^{(0)}$ is the bare coupling of
the model in the large-$N$ limit. We shall see in the next section,
where the auxiliary field will be eliminated, that at NLO in the 
large-$N$ expansion the bare coupling of the $O(N)$ model differs from
$\lambda_B,$ it turns out to be a combination of $\lambda_B$ and
$\delta g$ (see (\ref{Eq:cc_sep1}) and (\ref{Eq:cc_sep2})).

With the counterterm functional $\Delta\Gamma$ explicitly determined by
(\ref{Eq:Phi_ct-funct}), one can give now in a compact form the
functional introduced in (\ref{Gamma-funct}):
\bea
\Gamma_\textnormal{NLO}[\hat\alpha,v,G_\pi,{\cal G}]&=&
\frac{N}{2}(m_B^2-i\hat c\hat\alpha)v^2
+i\delta\kappa_1\hat\alpha+\frac{3N}{2\lambda_B}\hat\alpha^2
-\frac{i}{2}\int_k[\textnormal{Tr}\ln{\cal G}^{-1}(k)+(N-1)\ln G_\pi^{-1}(k)]
\nonumber\\
&&
-\frac{1}{2}\int_k \big[k^2-m_B^2+i\hat c\hat\alpha\big]\big[
(N-1) G_\pi(k)+G_{\sigma\sigma}^{(0)}(k)\big]
-iv\sqrt{\frac{\lambda}{3}}\int_k G_{\alpha\sigma}^{(0)}(k)
\nonumber\\
&&
+\frac{c}{2}\int_k G_{\alpha\alpha}^{(0)}(k)
-\frac{\lambda}{12}\int_k G_{\alpha\alpha}^{(0)}(k)\Pi(k),
\label{eq:sum-phi}
\eea
where $\lambda_B$ is defined in (\ref{Eq:lambdaB1}) and we introduced 
the following notations:
\be
m_B^2=m^2+\frac{1}{N}\delta m^2,\qquad 
\hat c=1+\frac{1}{N}\delta g, \qquad c=1+\delta \kappa_0,\qquad
\Pi(k)=-i\int_p G_\pi(p) G_\pi(k+p).
\label{Eq:mBhc}
\ee

\section{Elimination of the auxiliary field\label{sec:elim}}

In this section the ${\cal O}(N^0)$ accurate renormalized
functional will be established for the original formulation of
the $O(N)$ model by eliminating the auxiliary field $\hat\alpha$ and
the propagator components related to it (i.e. $G_{\alpha\sigma}$ and
$G_{\alpha\alpha}$).  In order to achieve this one substitutes into
(\ref{eq:sum-phi}) the expressions of $\hat\alpha,
G_{\alpha\sigma}^{(0)},$ and $G_{\alpha\alpha}^{(0)}$ as found from
their respective equations.

\subsection{Determination of $\Gamma_\textnormal{NLO}[v,G_\pi,G_\sigma]$}

For rewriting the terms depending on $G_{\alpha\alpha}^{(0)}$ and
$G_{\alpha\sigma}^{(0)}$ one exploits their representations which allow
the expression of the result fully in terms of $\Pi(k)$ and
$G_{\sigma\sigma}^{(0)}.$ The latter will be replaced with $G_\sigma,$ 
the exact sigma propagator of the $O(N)$ model.
In this way one finds
\bea
 -\frac{i}{2}\int_k\textnormal{Tr}\ln{\cal G}^{-1}(k)&=&
-\frac{i}{2}\int_k\ln\textnormal{det}{\cal G}^{-1}(k)=
-\frac{i}{2}\int_k\ln\left[\left(1-\frac{\lambda}{6}I_\pi^F(k)\right)
i G_\sigma^{-1}(k)\right]\nonumber\\
&=&
-\frac{i}{2}\int_k \ln G_\sigma^{-1}(k)
-\frac{i}{2}\int_k \ln\left(1-\frac{\lambda}{6 c}\Pi(k)\right)
-\frac{i}{2}\int_k \ln(i c) + {\cal O}\left(\frac{1}{N}\right).
\label{Eq:Phi_S1}
\eea
In going from the first to the second line above we used that 
to ${\cal O}(1/N)$ accuracy 
\be
-i\int_p G_\pi(p) G_\pi(k+p)\Big|_\textnormal{div}=
-i \int_p D_\pi(p) D_\pi(k+p)\Big|_\textnormal{div}= T_d^{(0)},
\label{Eq:Gaa4}
\ee 
and therefore one has
\be
1-\frac{\lambda}{6} I_\pi^F(k)=c-\frac{\lambda}{6} I_\pi(k)=
c-\frac{\lambda}{6} \Pi(k)+ {\cal O}\left(\frac{1}{N}\right),
\label{Eq:I_Pi_relation}
\ee
where the neglected ${\cal O}(1/N)$ contribution is finite.

Using (\ref{Eq:noZ1}) for $G_{\alpha\alpha}^{(0)}$ one writes the
following chain of expressions for the last term of (\ref{eq:sum-phi}):
\bea
-\frac{\lambda}{12}\int_k G_{\alpha\alpha}^{(0)}(k)\Pi(k)&=&
-\frac{c}{2}\int_k G_{\alpha\alpha}^{(0)}(k)+
\frac{1}{2}\int_k G_{\alpha\alpha}^{(0)}(k)
\left(c-\frac{\lambda}{6}\Pi(k)\right)
\nonumber\\
&=&
-\frac{c}{2}\int_k G_{\alpha\alpha}^{(0)}(k)
-\frac{\lambda}{6 c}v^2\int_k\frac{G_\sigma(k)}{1-\frac{\lambda}{6c}\Pi(k)}
-\frac{i}{2}\int_k 1+{\cal O}\left(\frac{1}{N}\right).
\label{Eq:Phi_S2}
\eea
In writing the second line above one again uses (\ref{Eq:I_Pi_relation}).
The first term on the r.h.s. above is canceled against the last but one term 
of (\ref{eq:sum-phi}). Finally, for the last term in the second line of  
(\ref{eq:sum-phi}) one uses (\ref{Eq:I_Pi_relation}) to write:
\be
\displaystyle
-i v\sqrt{\frac{\lambda}{3}}\int_k G_{\alpha\sigma}^{(0)}(k)=
\frac{\lambda}{3 c}v^2\int_k\frac{G_\sigma(k)}{1-\frac{\lambda}{6c}\Pi(k)}
+{\cal O}\left(\frac{1}{N}\right).
\label{Eq:Phi_S3}
\ee

As a short digression from our current task we mention that one could
proceed by further eliminating also the pion and sigma propagators
using their respective NLO and LO equations. Then one obtains the
renormalized version of the effective potential as function of
$\hat\alpha$ and $v$ studied in \cite{andersen04,andersen08}. A sketch
of this derivation is presented in \hbox{\ref{sec:Veff}} together with a
check of the renormalization of the saddle point equation coming from
this potential.

Now, we proceed instead with the elimination of $\hat\alpha$ from
(\ref{eq:sum-phi}) keeping the variables of the original theory, i.e.
$v,G_\pi,$ and $G_\sigma$.  Actually, the simplest way is to complete
to full square the functional depending quadratically on $\hat\alpha$
and then to use the saddle point equation $\delta
\Gamma/\delta\hat\alpha=0.$ Replacing $G_{\sigma\sigma}^{(0)}$ by the
exact propagator $G_\sigma$ of the $O(N)$ model one obtains from the
$\hat\alpha$-dependent part
\be
\frac{\lambda_B}{24 N}\left[
N\hat c v^2+\hat c \int_k\big[(N-1)G_\pi(k)+G_\sigma(k)\big]
-2\delta\kappa_1 \right]^2.
\label{Eq:Phi_Sa}
\ee

Putting together all above pieces one also uses that in view of 
(\ref{Eq:lambdaB3}) $\lambda/c=\lambda_B^{(0)}$ and obtains
\bea
\Gamma[v,G_\pi,G_\sigma]&=&
\frac{N}{2}\left(m_B^2-\frac{\lambda_B\hat c \delta\kappa_1}{3 N}\right)v^2
+N\frac{\lambda_B \hat c^2}{24} v^4
-\frac{i}{2}\int_k\big[(N-1)\ln G_\pi^{-1}(k)+\ln G_\sigma^{-1}(k)\big]
\nonumber\\
&&
-\frac{1}{2}\int_k \left[
k^2-m_B^2+\frac{\lambda_B\hat c \delta\kappa_1}{3 N}
-\frac{\lambda_B \hat c^2}{6} v^2 \right]
\big[(N-1) G_\pi(k)+G_\sigma(k)\big]
\nonumber\\
&&
+\frac{\lambda_B^{(0)}}{6}v^2\int_k\frac{G_\sigma(k)}
{1-\lambda_B^{(0)}\Pi(k)/6}
-\frac{i}{2}\int_k\ln\bigg(1-\frac{\lambda_B^{(0)}}{6}\Pi(k)\bigg)
\nonumber\\
&&
+\frac{\lambda_B \hat c^2}{24} (N-2)\left(\int_k G_\pi(k)\right)^2
+\frac{\lambda_B \hat c^2}{12} \int_k G_\pi(k) \int_p G_\sigma(p),
\label{eq:sum-phi2}
\eea
where we omitted terms of order ${\cal O}(1/N)$ and a divergent
constant $\sim \delta\kappa_1^2$ coming from (\ref{Eq:Phi_Sa}) as well
as the irrelevant divergent last but one terms of 
(\ref{Eq:Phi_S1}) and (\ref{Eq:Phi_S2}).

The bare couplings appearing in (\ref{eq:sum-phi2}) have to be
given with an accuracy corresponding to that of the functional 
(\ref{Gamma-funct}), therefore one writes the counterterms as a sum of
the LO and NLO contributions:
\begin{alignat}{4}
& \lambda_B \hat c^2=\lambda+\delta\lambda,\qquad
& &\delta\lambda =\delta\lambda^{(0)}+\frac{1}{N}\delta\lambda^{(1)},
& &
\nonumber \\ 
& m_B^2-\frac{\lambda_B\hat c \delta\kappa_1}{3 N}=m^2+\delta m^2, \qquad
& &\delta m^2 =\delta {m^2}^{(0)}+\frac{1}{N}\delta {m^2}^{(1)}.
& &
\label{Eq:cc_sep1}
\end{alignat}

With help of (\ref{Eq:lambdaB1}), (\ref{Eq:lambdaB2}), (\ref{Eq:lambdaB3}), 
(\ref{Eq:mBhc}) and the separation 
$\delta\kappa_1=N\delta\kappa_1^{(0)}+ \delta\kappa_1^{(1)},$
the counter-couplings above can be given in terms of the counter-couplings 
of the model in the auxiliary field formalism as
\begin{alignat}{4}
& \delta\lambda^{(0)}=\delta\lambda_\alpha^{(0)},\qquad
& & \delta\lambda^{(1)}=\delta\lambda_\alpha^{(1)}+
2\lambda_B^{(0)}\delta g,
& &
\nonumber \\
& \delta {m^2}^{(0)}=-\frac{1}{3}\lambda_B^{(0)}\delta\kappa_1^{(0)},\qquad
& & \delta {m^2}^{(1)}=\delta m^2-\frac{1}{3}\left[
\delta\lambda_\alpha^{(1)}\delta\kappa_1^{(0)}+\lambda_B^{(0)}
\left(\delta\kappa_1^{(1)}+\delta\kappa_1^{(0)}\delta g\right)
\right].
& &
\label{Eq:cc_sep2}
\end{alignat}

Using (\ref{Eq:cc_sep1}) in (\ref{eq:sum-phi2}) one obtains 
at NLO in the $1/N$ expansion the renormalized functional 
of the original $O(N)$ model, that is without the auxiliary field, 
in the following form:
\bea
\Gamma_\textnormal{NLO}[v,G_\pi,G_\sigma]&=&
-\frac{i}{2}\int_k \left[
(N-1)\left(\ln G_\pi^{-1}(k)+\mathscr{D}^{-1}_\pi(k) G_\pi(k)\right)
+\ln G_\sigma^{-1}(k)+\mathscr{D}^{-1}_\sigma(k) G_\sigma(k)
\right]
\nonumber\\
&+&\frac{N}{2} m^2 v^2+N\frac{\lambda}{24} v^4
+N\frac{\lambda}{24}\left(\int_k G_\pi(k)\right)^2
+\frac{\lambda}{12} \int_k G_\pi(k) \int_p G_\sigma(p)
\nonumber\\
&-&
\frac{\lambda_B^{(0)}}{6}v^2\int_k G_\sigma(k)
+\frac{\lambda_B^{(0)}}{6}v^2\int_k\frac{G_\sigma(k)}
{1-\lambda_B^{(0)} \Pi(k)/6}
\nonumber\\
&-&\frac{\lambda_B^{(0)}}{12}\left(\int_k G_\pi(k)\right)^2
-\frac{i}{2}\int_k\ln\bigg(1-\frac{\lambda_B^{(0)}}{6}\Pi(k)\bigg)
\nonumber\\
&+&
\frac{N}{2} \delta m^2 v^2+N\frac{\delta\lambda}{24} v^4
+\frac{\delta m^2}{2}\int_k\big[(N-1) G_\pi(k)+ G_\sigma(k)\big]
\nonumber\\
&+&\frac{\delta\lambda}{12}\left[
v^2\int_k\big[(N-1) G_\pi(k)+3 G_\sigma(k)\big]
+\frac{N}{2}\left[\int_k G_\pi(k)\right]^2
+\int_k G_\pi(k) \int_p G_\sigma(p)
\right],\ \ \ \ 
\label{Eq:Phi_ON}
\eea
where we introduced the usual tree-level propagators 
for the sigma and pion fields as
\be
i\mathscr{D}^{-1}_\sigma(k)=k^2-m^2-\frac{\lambda}{2} v^2,\qquad
i\mathscr{D}^{-1}_\pi(k)=k^2-m^2-\frac{\lambda}{6} v^2.
\ee

If one forgets about the counterterms, the expression in
(\ref{Eq:Phi_ON}) coincides with the 2PI effective potential obtained
in \cite{dominici93}. The terms above can be combined in a way which
makes explicit that in the large-$N$ expansion there are only two bare
couplings, namely $m^2+\delta m^2$ and $\lambda+\delta\lambda$. Would
one attempt a selfconsistent solution of the equations arising from
the variation of $\Gamma_\textnormal{NLO}$ this would not be true 
(see the concluding part of this subsection).  We emphasize
also that since the above functional is ${\cal O}(N^0)$
accurate, in terms involving the sigma propagator $G_\sigma$ one does
not need the NLO part of the counterterms, i.e.  $\delta m^{2^{(1)}}$
and $\delta\lambda^{(1)}.$ When differentiating with respect to
$G_\pi$ one should remember that also $\Pi(k)$ is a functional of
$G_\pi!$

The interpretation of the terms in (\ref{Eq:Phi_ON}), which makes explicit
the infinite series of diagrams summed up in the present treatment is as
follows. The first two and the last two lines represent the 2PI effective
potential of the $O(N)$ model at Hartree level of the truncation and at NLO
in its large-$N$ expansion. The remaining four terms represent the NLO
contribution of the 2PI vacuum diagrams beyond Hartree level. The
$v^2$-dependent part of these terms can be rewritten as
\bea
\frac{\lambda+\delta\lambda^{(0)}}{6} v^2 \int_k G_\sigma(k)
\frac{\big(\lambda+\delta\lambda^{(0)}\big)\Pi(k)/6}
{1-\big(\lambda+\delta\lambda^{(0)}\big) \Pi(k)/6}
=
\frac{\lambda+\delta\lambda^{(0)}}{6} v^2
\int_k G_\sigma(k)\sum_{n=1}^\infty
\left(\frac{\lambda+\delta\lambda^{(0)}}{6}\Pi(k)\right)^n.
\label{Eq:sum_1}
\eea
The $v^2$-independent part can also be written as a sum:
\bea
-\frac{i}{2}\left[\frac{\lambda+\delta\lambda^{(0)}}{6}\int_k \Pi(k)
+\int_k\ln\bigg(1-
\frac{\lambda+\delta\lambda^{(0)}}{6}\Pi(k)\bigg)
\right]=
\frac{i}{2}\int_k \sum_{n=2}^\infty\frac{1}{n}
\left(\frac{\lambda+\delta\lambda^{(0)}}{6}\Pi(k)\right)^n.
\label{Eq:sum_2}
\eea
It is easy to show that not considering counterterms, these terms correspond
to the two sets of diagrams given in Fig.~\ref{Fig:2sets}. Counterterm
diagrams corresponding to the $n=2$ term of the sum in (\ref{Eq:sum_1}) are
displayed in Fig.~\ref{Fig:ctset1}. Similar diagrams with different number
of pion bubbles can be drawn for the other terms appearing in the sums in 
(\ref{Eq:sum_1}) and (\ref{Eq:sum_2}). A direct way to obtain 
(\ref{Eq:Phi_ON}) consists of summing up all these diagrams with the 
associated combinatorial factors determined by the Feynman rules.

\begin{figure}[!t]
\centerline{ 
\includegraphics[keepaspectratio,width=0.67\textwidth,angle=0]{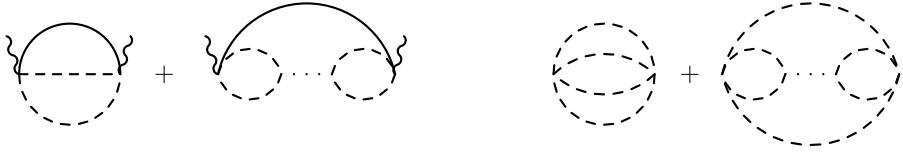}}
\caption{The two sets of vacuum diagrams which contribute beyond Hartree level 
and at NLO to the 2PI functional. Solid (dashed) line represents 
sigma (pion) propagator, while wiggly line represents the background $v.$ 
The dots indicate any number of pion bubbles.}
\label{Fig:2sets}
\end{figure}

\begin{figure}[!t]
\centerline{ 
\includegraphics[keepaspectratio,width=0.85\textwidth,angle=0]
{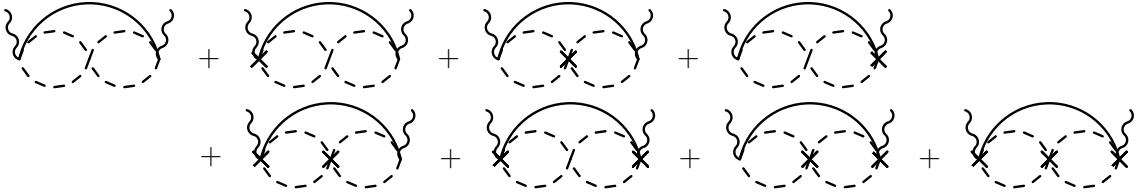}}
\caption{Diagrams corresponding to the $n=2$ term in the sum on the r.h.s. of
(\ref{Eq:sum_1}). The cross represents the counter-coupling
 $\delta\lambda^{(0)}.$}
\label{Fig:ctset1}
\end{figure}

An interesting question is how the counterterms of the last two lines of
(\ref{Eq:Phi_ON}) are related to the set of counterterms introduced in
\cite{berges05}. That structure was fixed by the possible $O(N)$
invariants of the model. Comparing the expressions presented in Eq.~(6)
of \cite{fejos08} with those of the last two lines of (\ref{Eq:Phi_ON}) 
one obtains the following unique relation for all of them:
\be
\delta\lambda_4=\delta\lambda_2^A=\delta\lambda_0^A=
\delta\lambda_2^B=\delta\lambda_0^B=\delta\lambda,\qquad
\delta m^2_2=\delta m^2_0=\delta m^2.
\ee
This result is actually expected on general grounds, since in the
strict $1/N$ expansion used in our approach different definitions of
the two- and four-point functions coincide. The above coincidence
represents rather a check of the correctness of our NLO renormalization.

\subsection{Renormalizability checks on $\Gamma_\textnormal{NLO}[v,G_\pi,G_\sigma]$}

The significance of (\ref{Eq:Phi_ON}) is that it displays all 
counterterms which guarantee the renormalizability of the resummation
of the perturbative series, a resummation induced by the large-$N$ expansion. 
The finiteness of the equation of
state and the self-energies obtained from its respective variations is
ensured automatically, since this feature is ``inherited'' from the
finiteness of the same quantities achieved in the formulation with the
auxiliary field. Still, it is an instructive exercise to check this
feature directly. Exploiting the structure of our previous analysis 
done in the auxiliary formulation of the model we shall show the 
finiteness of the equations directly obtained from (\ref{Eq:Phi_ON}). 

The inverse pion propagator at NLO in the $1/N$ expansion is given then by
\be
i G_\pi^{-1}(k)=k^2-M^2-\frac{\lambda}{3N}\Sigma_\pi^F(k),
\ee
where the nonlocal and local parts of the self-energy are:
\bea
\nonumber
\Sigma_\pi^F(k)&=&i\int_p\left[
-\frac{i}{1-\lambda \Pi_F(p)/6}
-\frac{\lambda v^2}{3} \frac{G_\sigma(p)}
{\left(1-\lambda \Pi_F(p)/6\right)^2}
\right]G_\pi(k+p)-\tilde T_\textnormal{div}(M^2),\\
M^2&=&m_B^2+\frac{\lambda_B}{6}\left(v^2+\int_k G_\pi(k)\right)
+\frac{\lambda_B}{6 N}\int_k \Big(G_\sigma(k)-G_\pi(k)\Big)
+\frac{\lambda}{3 N}\tilde T_\textnormal{div}(M^2).
\eea
For the nonlocal part we used that 
$6/\lambda_B^{(0)}-\Pi=6/\lambda-\Pi_F/6$ and $\tilde T_\textnormal{div}$ 
is given in (\ref{tilde-div}).

Since the local part has LO and NLO contributions one writes 
$M^2={M^2}^{(0)}+{M^2}^{(1)}/N$ and expands the pion 
propagator to ${\cal O}(1/N)$. With help of the integrals defined in 
(\ref{Eq:J_ints}) one obtains 
\bea
\nonumber
&&{M^2}^{(0)}=m^2+\delta {m^2}^{(0)}+\frac{\lambda_B^{(0)}}{6}
\left(v^2+\int_k D_\pi(k)\right),
\\
\nonumber
&&3 i \frac{{M^2}^{(1)}}{\lambda_B^{(0)}}\left(
1-\frac{\lambda_B^{(0)}}{6} I_\pi(0)\right)=
\frac{3i}{\lambda_B^{(0)}}\left[\delta {m^2}^{(1)}+
\frac{\lambda}{3}\tilde T_\textnormal{div}\big({M^2}^{(0)}\big)\right]
+\frac{i}{2}\int_k \Big(G_\sigma(k)-D_\pi(k)\Big)
\\
&&\qquad\ +\frac{\lambda}{6}\left[\tilde J({M^2}^{(0)})
-\tilde T_\textnormal{div}({M^2}^{(0)})\int_k D_\pi^2(k) \right]
-i\frac{\lambda^2}{18}v^2 J({M^2}^{(0)})
+\frac{i\delta\lambda^{(1)}}{2\lambda_B^{(0)}}
\left(v^2+\int_k D_\pi(k)\right),
\label{Eq:M2LO_NLO}
\eea
where in comparison to its definition given in (\ref{Eq:tree_prop}) 
$D_\pi$ depends now on ${M^2}^{(0)}.$ We used also that to leading order 
$\Pi(p)=I_\pi(p).$

The equation for ${M^2}^{(0)}$ in (\ref{Eq:M2LO_NLO}) is the usual 
gap equation at Hartree level of truncation of the effective action.
This was analyzed in \cite{fejos08} and the counterterms which
can be determine from this are $\delta \lambda^{(0)}$ and 
$\delta {m^2}^{(0)}$ given in (\ref{Eq:cc_sep2}).

Observing that the left hand side of the equation for ${M^2}^{(1)}$ 
in (\ref{Eq:M2LO_NLO}) is finite one obtains the following relation 
between counterterms and divergences:
\bea
-\frac{3i}{\lambda_B^{(0)}}\left[\delta {m^2}^{(1)}+
\frac{\lambda}{3}\tilde T_\textnormal{div}\big({M^2}^{(0)}\big)\right]
&=&\frac{\lambda}{6}\left[\tilde J_\textnormal{div}({M^2}^{(0)})
-\tilde T_\textnormal{div}({M^2}^{(0)})\int_k D_\pi^2(k) \right]
-i\frac{\lambda^2}{18}v^2 J_\textnormal{div}({M^2}^{(0)})
\nonumber\\
&&+\frac{i\delta\lambda^{(1)}}{2\lambda_B^{(0)}}
\left(v^2+\int_k D_\pi(k)\right)
+\frac{i}{2}\int_k 
\Big(G_\sigma(k)-D_\pi(k)\Big)\bigg|_\textnormal{div}.
\eea
Using the divergence analysis of \ref{sec:Div} one has all 
divergences and integrals expressed in terms of ${M^2}^{(0)}$ for 
which one can substitute its finite gap equation 
${M^2}^{(0)}=m^2+\lambda/6(v^2+T_\pi^F).$ Requiring the vanishing of 
the coefficient of $v^2+T_\pi^F$ determines $\delta\lambda^{(1)},$ 
while the remaining overall divergence determines 
$\delta {m^2}^{(1)}.$ Both are in accordance with (\ref{Eq:cc_sep2}).

The equation for the inverse sigma propagator obtained from 
(\ref{Eq:Phi_ON}) is
\be
i G_\sigma^{-1}(k)=k^2-{M^2}^{(0)}-\frac{\lambda_B^{(0)}}{3}v^2 
\frac{1}{1-\lambda_B^{(0)} I_\pi(k)/6},
\ee
which is finite, since ${M^2}^{(0)}$ is finite and 
$6/\lambda_B^{(0)}-I_\pi=6/\lambda-I_\pi^F/6.$

Using the relation between the LO pion and sigma propagators one can show 
that the derivative of (\ref{Eq:Phi_ON}) with respect to $v$ reads as
\be
\frac{\delta \Gamma_\textnormal{NLO}}{\delta v}[v,G_\pi,G_\sigma]=
-N v i G_\pi^{-1}(k=0),
\ee
which is also finite, since we showed that $G_\pi^{-1}(k)$ is finite.
It also displays the validity of Goldstone's theorem.

We close this part by mentioning that the LO sigma propagator
equation, the NLO pion propagator equation and the equation of state
derived from (\ref{Eq:Phi_ON}) can be obtained also with the
Dyson-Schwinger formalism of Section~2. Concerning these equations,
the large-$N$ expansion closes the hierarchy of DS equations at the
level of complete LO renormalized $\Gamma_{\pi\pi\sigma}$ and
$\Gamma_{\pi\pi\pi\pi}$ vertex functions, which include also one-loop
pion corrections.  Some details can be found in \cite{patkos06} for
$G_\sigma$ and $\Gamma_{\pi\pi\sigma}$ (see also \cite{korpa} for the
relation between the truncation of the Dyson-Schwinger equations and
the $1/N$ expansion).  This means that our investigation implies also
the renormalizability of the Dyson-Schwinger equations at NLO in the
large-$N$ expansion.

\section{Conclusions}

We studied the renormalizability of the $O(N)$ model at
next-to-leading order in the large-$N$ expansion, at zero
temperature. We constructed the ${\cal O}(N^0)$ counterterm 
functional of the model in auxiliary field formalism by 
studying the renormalization of the derivatives of a generalized 
effective action functional with respect to its variables. 
Expanding the propagator equation of the
pion to ${\cal O}(1/N)$ in the large-$N$ expansion we showed that the
renormalization can be achieved for arbitrary values of the background
and auxiliary field, in a way that respects the internal symmetry of
the model (e.g. Goldstone's theorem). This can be expected on
theoretical grounds, since divergences are determined only by the
asymptotic behavior of the propagators and because any consistent
resummation of the perturbation theory should resum also the
counterterm diagrams associated to the perturbative series.  Although,
one can anticipate the consistency of the auxiliary field technique
and of the large-$N$ expansion in dealing with perturbative series,
the difficulty we face when trying to infer the renormalizability of
the model in a given approximation from the fact that the model is
perturbatively renormalizable is that one cannot easily keep track of
what partial series of counterterm diagrams is actually resummed at a
given order of the large-$N$ expansion. In consequence, the actual
analytic check of the renormalization of a given approximation is
unavoidable.

The elimination of the auxiliary field and the related propagators,
while keeping the dynamical sigma and pion propagators, makes
transparent the classes of diagrams containing also counterterms which
are resummed in the original $O(N)$ theory at NLO of the large-$N$
expansion. The explicit form of the counterterms is given here for the
first time for the theory using auxiliary field and also for the
original formulation. In the original theory the action functional
contains only two counterterms, a coupling and a mass counterterm,
both having LO and NLO parts. The propagator equations and the
equation of state derived from that coincides with the 1PI
Schwinger-Dyson equations closed at the complete LO
$\Gamma_{\pi\pi\sigma}$ and $\Gamma_{\pi\pi\pi\pi}$ vertex functions,
which includes one-loop pion corrections.

The two examples we worked out
(e.g. $\Gamma_\textnormal{NLO}[v,G_\pi,G_\sigma]$ and
$\Gamma_\textnormal{NLO}[\hat\alpha,v]$) demonstrate that the
renormalizability of the broadest action functional
$\Gamma[\hat\alpha,v,G_\pi,{\cal G}]$ implies the renormalizability of
the functionals arising after the elimination of some subset
of the variables. This result obtained at $T=0$ makes us confident
that the renormalization goes through for nonzero temperature as well
and that the counterterm functional determined here will prove helpful
for phenomenological studies in the $O(N)$ model. A study of the
renormalization scale invariance at NLO in different formulations
would be helpful for such applications.  The method developed here can
be used also to models with more complicated global symmetries.

\begin{acknowledgments}
The authors benefited from discussions with Prof. Peter Sz{\'e}pfalusy.
This work is supported by the Hungarian Research Fund under contracts
Nos. T046129 and T068108. Zs. Sz. was supported by OTKA Postdoctoral Grant No. 
PD 050015.
\end{acknowledgments}

\appendix
  
\section{Detailed analysis of the divergences \label{sec:Div}}

\subsection*{Notations}
Following the spirit of Ref.~\cite{patkos08} we will expand the
propagators around appropriately chosen infrared safe auxiliary
propagators. Since at NLO the asymptotics is determined basically by
integrals of the LO propagator $D_\pi(p)$ and of the propagator
$\tilde G(p)$ given in (\ref{Eq:Gt}) which incorporates the effect of
the resummation of the chain of pion bubbles, one needs two auxiliary
propagators. The first one is
\be
G_0(p)=\frac{i}{p^2-M_0^2},
\ee
where $M_0$ is an arbitrary mass scale. With this propagator one 
defines the quadratically divergent integral
\be
T_d^{(2)}=\int_p G_0(p),
\label{Eq:Td2}
\ee
and the following one-loop bubble integral
\be
I_0(p)=-i \int_k G_0(k) G_0(k+p)=T_d^{(0)}+I_0^F(p).
\ee
The logarithmically divergent part of the integral above is defined as
\be
T_d^{(0)}=-i \int_k G_0^2(k).
\label{Eq:Td0}
\ee
The finite part behaves asymptotically as
$I_0^F(p)\sim\ln\frac{p^2}{M_0^2}-2-i\pi+{\cal O}\left(p^{-2}\ln p\right),$
and defines together with $G_0(p)$ the second
auxiliary propagator:
\be
G_a(p)=\frac{i}{(p^2-M_0^2)(1-\lambda I_0^F(p)/6)}.
\ee
The integrals involving only combinations of $G_0(p),$ $G_a(p)$ and
$I_0^F(p)$  will be fully included in the counterterms. 
With help of the propagator $G_0(p)$ one can separate the quadratic and
logarithmic divergence of the tadpole integral defined with the tree-level
propagator given in (\ref{Eq:tree_prop}). At zero temperature one has:
\be
\int_k D_\pi(k)=
T_d^{(2)}+\left(M^2-M_0^2\right)T_d^{(0)}+T_\pi^F,\qquad
T_\pi^F=\frac{1}{16\pi^2}\left(
M^2\ln\frac{M^2}{M_0^2}-M^2+M_0^2\right). 
\label{Eq:tadpole}
\ee

\subsection*{Absence of wave function renormalization in the NLO pion propagator}

The only term which on dimensional grounds could produce a momentum-dependent
divergence in the expression of the NLO pion propagator 
(\ref{Eq:Gp_NLO}) is the first one on the r.h.s. of (\ref{Eq:noZ1}).
It gives the integral
\be
\int_p \frac{D_\pi(p+k)}{1-\lambda I_\pi^F(p)/6}.
\label{Eq:noZ2}
\ee
To study this integral one uses the following expansion
\be
-i D_\pi(p+k)=\frac{1}{(p+k)^2-M^2}=\frac{1}{p^2-M^2}
+\frac{1}{p^2-M^2}\sum_{n=1}^\infty
\left(-\frac{k^2+2 p\cdot k}{p^2-M^2}\right)^n.
\label{Eq:noZ3}
\ee
In order to prove that there is no infinite wave function renormalization it
is enough to look at the appearance of $k^2$ in the numerator, that is
at terms $n=1,2$ in the sum. Keeping only terms up to and including 
${\cal O}(1/p^4),$ but throwing away those which vanish upon
symmetrical integration
(note that $I_\pi^F(p)$ depends on $p^2$) amounts to the following 
replacement at the level of the integrand in (\ref{Eq:noZ2})~:
\be
\frac{1}{(p+k)^2-M^2}\longrightarrow \frac{1}{p^2-M^2}+
\frac{4 (p\cdot k)^2 - k^2 p^2}{(p^2-M^2)^3}.
\label{Eq:noZ4}
\ee
The second term above gives vanishing contribution in
(\ref{Eq:noZ2}) due to the following
property which holds for any integrable function $f(p^2)$ upon the use of a Lorentz-invariant
regulator: 
\[
\int d^4 p\, p_\mu p_\nu f(p^2)=\frac{g_{\mu\nu}}{4}
\int d^4 p\, p^2 f(p^2).
\]
The first term on the r.h.s. of (\ref{Eq:noZ4}) gives the momentum 
independent divergence denoted with 
$\tilde T_\textnormal{div}$ in (\ref{pion-p0}).

\subsection*{Momentum independent overall divergence of the NLO pion propagator}

To find the momentum independent divergence of the NLO pion propagator
one starts with (\ref{pion-p0}) and takes into account that
the one-loop bubble integral behaves logarithmically for large momentum. 

The divergence of the one-loop bubble integral is chosen to be $T_d^{(0)}$ 
given in (\ref{Eq:Td0}). Then, one has 
\be
I_\pi(p)=-i\int_k D_\pi(k) D_\pi(k+p)=T_d^{(0)}+I_\pi^F(p),\qquad
I_\pi^F(p)=\frac{1}{16\pi^2}\left[-2+\ln\frac{M^2}{M_0^2}+L(p,M)\right].
\label{Eq:mid1}
\ee
Here, the momentum dependent function $L(p^2,M)$ which determines the 
finite part $I_\pi^F(p)$ can be found in Eq.~(8) of Ref.~\cite{patkos02b}. 
From $\underset{p\to 0}{\lim} L(p,M)=2,$ it results that
\be
I_\pi(p=0)=-i\int_k D_\pi^2(k)=
T_d^{(0)}+I_\pi^F(p=0),\qquad
I_\pi^F(p=0)=\frac{1}{16\pi^2}\ln\frac{M^2}{M_0^2}.
\label{Eq:Bp0}
\ee
Since $I_0^F(p)$ has exactly the same form as 
$I_\pi^F(p),$ but with $M^2$ replaced by $M_0^2,$ for large $p^2$ one has
\be
I_\pi^F(p)-I_0^F(p)=\frac{1}{8\pi^2}\frac{M_0^2-M^2}{p^2}\ln\frac{p^2}{M_0^2}
+{\cal O}\left(\frac{1}{p^2}\right).
\label{Eq:mid2a}
\ee
In the asymptotic momentum region the above expression allows us to write 
\bea
\nonumber
\frac{1}{1-\lambda I_\pi^F(p)/6}&=&\frac{1}{1-\lambda I_0^F(p)/6}+
\frac{\lambda}{6}\frac{I_\pi^F(p)-I_0^F(p)}{(1-\lambda I_0^F(p)/6)^2}
+{\cal O}\left(\frac{1}{p^4 \ln p}\right)\\
&=&\frac{1}{1-\lambda I_0^F(p)/6}+
\frac{\lambda}{3}\frac{M_0^2-M^2}{p^2-M_0^2}
\frac{I_0^F(p)}{(1-\lambda I_0^F(p)/6)^2}+{\cal O}
\left(\frac{1}{p^2\ln^2 p}\right),
\label{Eq:mid2b}
\eea
where in the last line we used
$I_0^F(p)\sim\ln\frac{p^2}{M_0^2}-2-i\pi+{\cal O}\left(p^{-2} \ln p\right).$
The neglected terms give finite contribution in the last 
integral of (\ref{pion-p0}). Using there (\ref{Eq:mid2b}) and  
\bea
\frac{1}{p^2-M^2}=\frac{1}{p^2-M_0^2}+\frac{M^2-M_0^2}{(p^2-M_0^2)^2}+
{\cal O}\left(\frac{1}{p^6}\right),
\label{Eq:mid3}
\eea
one obtains
\be
\tilde T_\textnormal{div}(M^2)=T_a^{(2)}-(M^2-M_0^2) i\int_p G_a^2(p)
\left(1-\frac{\lambda}{6}I_0^F(p)\right)\bigg|_\textnormal{div}
-\frac{\lambda}{3} (M^2-M_0^2) T_a^{(I)},
\label{Eq:mid4}
\ee
where the following divergent integrals were defined
\be
T_a^{(2)}=\int_p G_a(p),\qquad T_a^{(I)}=-i \int_p G_a^2(p) I_0^F(p).
\label{Eq:Ta2I}
\ee
Using in the remaining integral of (\ref{Eq:mid4})
that $\int_p G_a^2(p)$ is finite, one obtains for 
$\tilde T_\textnormal{div}$ the expression given in (\ref{tilde-div}).

\subsection*{Analysis of the divergences of the saddle point equation 
(\ref{Eq:Act1}) }

We start with the differences of tadpoles involving LO sigma and pion 
propagators. Using (\ref{Eq:LO_relation}) iteratively one finds
\be
\left[\frac{i}{2}\int_k\left(G_{\sigma\sigma}^{(0)}(k)
-D_\pi(k)\right)\right]\bigg|_\textnormal{div}=
\frac{\lambda}{6}v^2 \int_k \frac{D_\pi^2(k)}{1-\lambda I_\pi^F(k)/6}
\bigg|_\textnormal{div}.
\label{Eq:v2_div1}
\ee
In view of (\ref{Eq:mid2b}) one can replace $I_\pi^F(k)$ with
$I_0^F(k)$ in the denominator above. Then using (\ref{Eq:mid3}) one
obtains
\be
\left[\frac{i}{2}\int_k\left(G_{\sigma\sigma}^{(0)}(k)
-D_\pi(k)\right)\right]\bigg|_\textnormal{div}=
\frac{\lambda}{6}v^2 
\int_k G_a^2(k)
\left(1-\frac{\lambda}{6}I_0^F(k)\right)\bigg|_\textnormal{div}=
-i\frac{\lambda^2}{36}v^2T_a^{(I)}.
\label{Eq:v2_div2}
\ee

Next, we investigate the second double integral given in (\ref{Eq:J_ints}). 
Changing the order of integration one uses (\ref{Eq:LO_relation}) and the
following relation which can be derived from (\ref{Eq:mid1})
\be
\int_k D_\pi^2(k) D_\pi(p+k)=-\frac{1}{2}
\frac{d }{d M^2} I_\pi(p)=\frac{1}{p^2-4 M^2}\left[
I_\pi^F(p)-\frac{1}{16\pi^2}\ln\frac{M^2}{M_0^2}+\frac{1}{8\pi^2}
\right],
\label{Eq:v2_div3}
\ee
to find 
\be
J_\textnormal{div}(M^2)=
\int_p \frac{D_\pi(p)}{(1-\lambda I_\pi^F(p)/6)^2}
\frac{I_\pi^F(p)}{p^2-4 M^2}\Bigg|_\textnormal{div}
=-i\int_p G_a^2(p) I_0^F(p)=T_a^{(I)}.
\label{Eq:v2_div4}
\ee
To obtain the second equality above we replaced $I_\pi^F$ with $I_0^F$
in view of (\ref{Eq:mid2a}) and used (\ref{Eq:mid3}).

The first double integral given in (\ref{Eq:J_ints}) contains an 
overall divergence as well as subdivergences. By changing the order of
integration and using (\ref{Eq:v2_div3}), one has
\be
\tilde J_\textnormal{div}(M^2)=
\frac{6}{\lambda}\int_p \frac{-1}{p^2-4M^2}\Bigg|_\textnormal{div}
+\left(\frac{6}{\lambda}+\frac{1}{8\pi^2}-\frac{1}{16\pi^2}
\ln\frac{M^2}{M_0^2} \right)
\int_p \frac{1}{p^2-4M^2}\frac{1}{1-\lambda I_\pi^F(p)/6}
\Bigg|_\textnormal{div},\ 
\label{Eq:odiv1}
\ee
where we have separated a divergence independent of $I_\pi^F(p),$
which by using a relation similar to (\ref{Eq:mid3}) can be expressed
as a linear combination of $T_d^{(2)}$ and $T_d^{(0)}$. 

Since the form of the second integral in (\ref{Eq:odiv1}) is similar to
the last one in (\ref{pion-p0}),  the calculation follows very closely 
the determination of $\tilde T_\textnormal{div}$. 
Using (\ref{tilde-div}) and (\ref{Eq:Bp0}) the result is
\bea
\nonumber
\frac{\lambda}{6}\tilde J_\textnormal{div}(M^2)&=&
i T_d^{(2)}+(4M^2-M_0^2)i T_d^{(0)}\\
&&
+i\left(\frac{\lambda}{2} M^2 T_a^{(I)}-\tilde T_\textnormal{div}(M^2)\right)
\left(1+\frac{\lambda}{6}T_d^{(0)}+\frac{\lambda}{48\pi^2}-
\frac{\lambda}{6}I_\pi(p=0)\right).
\label{Eq:odiv2}
\eea

\subsection*{Cut-off dependence of the
divergent integrals}
It is instructive to evaluate explicitly the cut-off dependence of the
divergent integrals denoted with $T_d^{(0)},T_d^{(2)},T_a^{(0)}$ and
$T_a^{(2)}.$ This can be also of some practical interest when one
proceeds to the numerical solution of the renormalized equations.
Going to Euclidean space with $p_0\to i p_E^0$ and using a 4d cut-off
$\Lambda$ the first two integrals defined in (\ref{Eq:Td0}) and
(\ref{Eq:Td2}) can be done analytically:
\[
T_d^{(0)}=\frac{1}{16\pi^2}\left[
\ln\left(\frac{\Lambda^2}{M_0^2}+1\right)
-\frac{\Lambda^2}{\Lambda^2+M_0^2}
\right],\qquad
T_d^{(2)}=\frac{1}{16\pi^2}\left[
\Lambda^2+M_0^2\ln\frac{M_0^2}{\Lambda^2+M_0^2}
\right].
\]
For the other two integrals defined in (\ref{Eq:Ta2I})
we limit ourselves to an asymptotic analysis 
and expand the integrand for large $k$. 
Exploiting the freedom to omit those contributions to $T_a^{(I)}$
which are formally finite for $\Lambda\to \infty,$ 
we choose the scheme in which
\be
T_a^{(I)}=-\frac{2}{(4\pi)^4}\int^\Lambda \frac{d k}{k}
\frac{\ln(k^2/M_0^2)}{(1+2 a -a \ln(k^2/M_0^2))^2}.
\label{Eq:TaI_calc}
\ee
Here, we introduced $a=\lambda/(96\pi^2).$ One notices that
for $k=M_0\exp(1+48\pi^2/\lambda)$ the denominator of the integral
above vanishes. To avoid this non-integrable singularity to occur in
the range of integration, that is for $k<\Lambda,$ one needs
$\Lambda<\Lambda_\textnormal{max}=M_0\exp(1+48\pi^2/\lambda).$ That
means that for $\lambda\ne 0$ the cut-off cannot be sent to infinity,
there is a maximal value for it, which reflects the triviality
of the theory.

Performing the integral in (\ref{Eq:TaI_calc}), and obeying this 
restriction, one finds
\[
T_a^{(I)}=-\frac{36}{\lambda^2}\ln\left(
-\frac{\lambda}{96\pi^2}\ln\frac{\Lambda^2}{M_0^2}+1+\frac{\lambda}{48\pi^2}
\right).
\]
With the same strategy one can choose
\[
T_a^{(2)}=\frac{3 M_0^2}{8\pi^2\lambda}
\left[-e^{2+96\pi^2/\lambda}\textnormal{Ei}\left(
\ln\frac{\Lambda^2}{M_0^2}-2-\frac{96\pi^2}{\lambda}\right)
+3 \ln\left(
-\frac{\lambda}{96\pi^2}\ln\frac{\Lambda^2}{M_0^2}+1+\frac{\lambda}{48\pi^2}
\right)
\right].
\]

\section{Derivation of the effective potential 
$V_\textnormal{NLO}[\hat\alpha,v]$ \label{sec:Veff} }

One starts from (\ref{eq:sum-phi}) and after using (\ref{Eq:Phi_S2}) and 
(\ref{Eq:Phi_S3}) one finds that in view of (\ref{Eq:LO_relation})
$G_{\sigma\sigma}^{(0)}\equiv G_\sigma$ drops out from the functional, 
which now becomes
\bea
\Gamma[\hat\alpha,v,G_\pi]&=&\frac{N}{2}(m_B^2-i\hat c\hat\alpha)v^2
+i\delta\kappa_1\hat\alpha+\frac{3N}{2\lambda_B}\hat\alpha^2
-\frac{i}{2}(N-1)\int_k
\left(\ln G^{-1}_\pi(k) + D^{-1}_\pi(k)G_\pi(k)\right)
\nonumber
\\
&&-\frac{i}{2}\int_k \ln\bigg[\big(k^2-m^2+i\hat\alpha\big)
\bigg(1-\frac{\lambda_B^{(0)}}{6}\Pi(k)\bigg)
-\frac{\lambda_B^{(0)}}{3} v^2\bigg].
\label{Eq:V2a}
\eea
Here, the last term comes from (\ref{Eq:Phi_S1}).
Next, one uses (\ref{pi-prop-1}), (\ref{tilde-div}), and the definitions in 
(\ref{Eq:mBhc}) to write the inverse pion propagator as 
\be
i G_\pi^{-1}(k)=k^2-m_B^2+i\hat c\hat\alpha-\frac{\lambda}{3 N} \Sigma_\pi(k),
\ee
where $\Sigma_\pi(k)$ is given by the integral of (\ref{pi-prop-1})
calculated with the expression $G_{\alpha\alpha}^{(0)}$ taken from
(\ref{Eq:noZ1}). Using this propagator in the first integral of 
(\ref{Eq:V2a}), one easily sees that when expanding it to ${\cal O}(1/N)$ 
the contribution of the self-energy drops out and we are left with
\bea
\Gamma_\textnormal{NLO}[\hat\alpha,v]&=&\frac{N}{2}(m_B^2-i\hat c\hat\alpha)v^2
+i\delta\kappa_1\hat\alpha+\frac{3N}{2\lambda_B}\hat\alpha^2
-N\frac{i}{2}\int_k \ln\big(k^2-m_B^2+i\hat c\hat\alpha\big)
\nonumber
\\
&&-\frac{i}{2}\int_k \ln\bigg(1-\frac{\lambda_B^{(0)}}{6}\Pi(k)
-\frac{\lambda_B^{(0)}}{3} v^2\frac{1}{k^2-m^2+i\hat\alpha}\bigg).
\label{Eq:V2b}
\eea

The radiative part of this functional has exactly the same form as the
effective potential given in \cite{root74,andersen04,andersen08}.  The
difference in the classical part corresponds to slightly different
ways of introducing the auxiliary field. More important is that the
authors of \cite{andersen04,andersen08} restrict their counterterm
functional only to terms proportional to pieces of the Lagrangian
which are present already in the original formulation of the model. In
their form of introducing the auxiliary field this restricts
the counterterms to those proportional to $\hat\alpha$ and
$\hat\alpha^2$. By allowing all independent counterterms to appear 
which have dimension less than or equal to 4 in the auxiliary field formulation 
one might expect to have enough flexibility to ensure the 
renormalisibility in arbitrary background 
\footnote{We thank J.O. Andersen and T. Brauner for discussions 
eventually permitting to localize the root of the apparent 
non-renormalisability of their treatment. }.

 In order to demonstrate that 
(\ref{Eq:V2b}) contains all the NLO counterterms, we sketch the 
renormalization of the SPE obtained by differentiating (\ref{Eq:V2b}) 
with respect to $\hat\alpha.$ Using that $\Pi,$ to be taken only at LO 
in $1/N$ expansion, depends on $\hat\alpha$ through $D_\pi$ defined 
in (\ref{Eq:tree_prop}), one obtains 
\bea
0&=&\frac{3N}{\lambda_B}\hat\alpha-i\delta\kappa_1
-i\frac{N}{2}\left(v^2+\int_k\frac{i}{k^2-m_B^2+i\hat c\hat\alpha}\right)
\nonumber
\\
&&-\frac{i}{2}\int_k\big(G_{\sigma\sigma}^{(0)}(k)-D_\pi(k)\big)
-\frac{\lambda}{6} \tilde J(M^2)+i\frac{\lambda^2}{18}v^2 J(M^2).
\label{Eq:Asep1}
\eea
Here, we recognized the appearance of the expression of 
$G_{\sigma\sigma}^{(0)}$ which can be read from (\ref{Eq:G_LO_matrix}).
We used the relation (\ref{Eq:LO_relation}) 
and for the last two terms also (\ref{Eq:I_Pi_relation}) and 
(\ref{Eq:J_ints}).

All we have to do is to establish the connection between (\ref{Eq:Asep1})
and (\ref{counter-spe-div}), the latter being already renormalized. The last
three terms of (\ref{Eq:Asep1}) can be found in
(\ref{counter-spe-div}), if in that equation one takes into account
(\ref{pion-tadpole}), so we have to work only on the first three terms
of (\ref{Eq:Asep1}). Using the definition of the couplings, 
one expands them to ${\cal O}(1/N)$ and obtains
\bea
i\frac{N}{2}
\left(v^2+\int_k\frac{i}{k^2-m_B^2+i\hat c\hat\alpha}\right)&=&
i\left(\frac{N}{2}+\frac{\lambda^2}{12}T_a^{(I)}\right)
\left(v^2+\int_k D_\pi(k)\right)
-\frac{\lambda}{6}\tilde T_\textnormal{div}(M^2)\int_k D_\pi^2(k),
\nonumber
\\
\frac{3N}{\lambda_B}\hat\alpha-i\delta\kappa_1&=&
\frac{3N}{\lambda}\hat\alpha
+i\frac{N}{2}\left[T_d^{(2)}+(M^2-M_0^2)T_d^{(0)}\right]
+i \delta\kappa_1^{(1)}+2\delta\kappa_2^{(1)}\hat\alpha.
\eea
Since the last two terms of the second equality 
above coincide with
$\delta\Delta\Gamma^{\alpha,0}(\hat\alpha)/\delta\hat\alpha$ of 
(\ref{counter-spe-div}), the equivalence between (\ref{Eq:Asep1}) and
(\ref{counter-spe-div}) is demonstrated.

}

\end{document}